# Modeling meso-scale energy localization in shocked HMX, Part II: training machine-learned surrogate models for void shape and void-void interaction effects

S. Roy[1], N. K. Rai[1], O. Sen[1], D. B. Hardin[2], A. S. Diggs[2] and H.S. Udaykumar[1*]

[1]*Mechanical Engineering, The University of Iowa, Iowa City, Iowa 52242, USA*

[2]*Air Force Research Laboratory, Munitions Directorate (AFRL/RW), Eglin, Florida 32542, USA*

Abstract

Surrogate models for hotspot ignition and growth rates were presented in Part I, where the hotspots were formed by the collapse of single cylindrical voids. Such isolated cylindrical voids are idealizations of the void morphology in real meso-structures. This paper therefore investigates the effect of non-cylindrical void shapes and void-void interactions on hotspot ignition and growth. Surrogate models capturing these effects are constructed using a Bayesian Kriging approach. The training data for machine learning the surrogates are derived from reactive void collapse simulations spanning the parameter space of void aspect ratio (AR), void orientation ($\theta$), and void fraction ($\phi$). The resulting surrogate models portray strong dependence of the ignition and growth rates on void aspect ratio and orientation, particularly when they are oriented at acute angles with respect to the imposed shock. The surrogate models for void interaction effects show significant changes in hotspot ignition and growth rates as the void fraction increases. The paper elucidates the physics of hotspot evolution in void fields due to the creation and interaction of multiple hotspots. The results from this work will be useful not only for constructing meso-informed macro-scale models of HMX, but also for understanding the physics of void-void interactions and sensitivity due to void shape and orientation.

## 1 INTRODUCTION

Closure terms are required in macroscale models of the shock response of heterogeneous energetic (HE) materials to capture the effects of sub-grid meso-scale thermophysical phenomena [1]. In calculations of shock to detonation transition (SDT) in HEs, these terms reflect energy localization at hotspots caused by void collapse or other mechanisms. In Sen et al. [2], a meso-informed multiscale model, MES-IG, was developed in the ignition and growth framework; machine learned surrogate models were used to connect the meso- and macro-scale response of an HE. The rate of energy localization due to hotspots was transmitted through the surrogate models to close the macro-scale equations [2]. These surrogate models supplied the rates of hotspot ignition ($\dot{F}_{ignition}$) and growth ($\dot{F}_{growth}$) as a function of the meso-structural parameters (void size) and loading (shock pressure and pulse duration). Part I [3] and this work together describe the techniques to construct surrogate models. While Part I [3] constructed the surrogate model for energy localization due to collapse of isolated cylindrical voids, in this work the effects of void shape variations and void-void interactions are quantified.

* Corresponding author email : ush@engineering.uiowa.edu

As discussed in a previous paper by Sen et al. [2], several loading and meso-structural parameters govern the ignition and growth functions, which can be modeled as:

$$\dot{F}_{ignition} = \dot{F}^{scv}_{ignition}(P_s, \tau_s, D_{void}) * f^{shape}_{ignition}(AR, \theta) * f^{v-v}_{ignition}(\phi) \tag{1}$$

and

$$\dot{F}_{growth} = \dot{F}^{scv}_{growth}(P_s, \tau_s, D_{void}) * f^{shape}_{growth}(AR, \theta) * f^{v-v}_{growth}(\phi) \tag{2}$$

Here $\dot{F}^{scv}_{ignition}$ and $\dot{F}^{scv}_{growth}$ represent the effect of loading (shock pressure $P_s$, shock pulse width $\tau_s$) and void size ($D_{void}$) on *single cylindrical voids (scv)*. Surrogate models for the response functions $\dot{F}^{scv}_{ignition}$ and $\dot{F}^{scv}_{growth}$ were constructed in Part I [3]. The focus of this paper is to build metamodels for the functions $f^{shape}_{ignition}(AR, \theta), f^{shape}_{growth}(AR, \theta), f^{v-v}_{ignition}(\phi), f^{v-v}_{growth}(\phi)$, which represent the effects of void shapes (parameterized by aspect ratio AR and orientation $\theta$) and void-void interactions in a void field (parameterized by void fraction $\phi$). The importance of capturing void shape and interaction effects, for predicting the response of an HE, are outlined below.

## 1.1 Importance of void shape and void-void interactions in energy localization

A typical meso-structure for pressed HMX [4] is shown in Figure 1 [5]. Clearly, the voids in the meso-structure are far from cylindrical in shape and are also surrounded by other voids. Thus, the ignition and growth rates of a hotspot may not be adequately represented by $\dot{F}^{scv}_{ignition}$ and $\dot{F}^{scv}_{growth}$. In fact, the deviations from cylindrical shape of voids, measured by the aspect ratio ($AR$) and orientation $\theta$, have been shown [6, 7] to have strong effects on void collapse dynamics and the resulting hotspot intensity (size and temperature). Simulations have also shown that void-void interactions can lead to significant differences in intensity of void collapse and the resulting hotspots in real meso-structures [8].

While the phenomenon of void collapse has been extensively studied via experiments [9-11] and numerical simulations [6, 8, 12-18], the focus has primarily been on ideal-shaped voids such as cylindrical, elliptical, conical, or spherical voids. There have only been a few published papers that have studied deviations from cylindrical shape. An early work is Tran et al. [13], where inert calculations were performed on spherical and triangular/conical voids in HMX. They observed that the spherical voids produced a lower intensity hotspot when compared to the triangular void. Also, the triangular void with its apex pointed towards the shock wave generated a low intensity hotspot when compared to a triangle with its base pointing towards the incident shock. In more recent work, Levesque and Vitello [6] analyzed the effect of void morphology on the hotspot temperature for TATB using inert meso-scale simulations. They showed that elliptical and cylindrical voids aligned with the shock along their longest dimension (i.e. along the major axis for the elliptical void) produced higher intensity hotspots than a spherical void for the same shock strength and void fraction. Recently, reactive meso-scale simulations have also been employed to study the effect of void morphology on the sensitivity of energetic materials. Kapila et al. [16] performed reactive simulations on three different void shapes: spherical, prolate spheroid (long void) and oblate spheroid (tall void) in HMX. Among the three void shapes, the oblate spheroid was found to generate higher temperature hotspots that can lead to detonation. The recent work by Springer et al. [17] analyzed the effect of void morphology on the shock initiation and reaction rate in HMX. It was again shown that elliptical and conical voids of high aspect ratio led to increased sensitivity. Additionally, the effect of void morphology is shown to be dominant in the low shock pressure loading situations. In recent work, 3D reactive void collapse simulations on non-spherical voids were performed by Rai et al. [19]. They show that, for favorable void aspect ratios and orientations, the hotspot size, temperature and the resulting growth rate are enhanced significantly when compared with a corresponding idealized cylindrical void. In fact, the amplification of hotspot intensity is even higher for three-dimensional voids [19] due to 3-dimensional baroclinic instabilities leading to enhanced mixing of reacted and unreacted material. The above studies show that closure models for energy localization at the meso-scale must take into account the effects of void shape.

Voids in HEs are seldom found in isolation; the collapse of one void can affect the intensity of collapse of neighboring voids. This void-void interaction effect arises due to blast waves emanating from the point of collapse of one void or secondary waves reflected from neighboring voids. These effects were studied in the case of void collapse in an inert material in previous work [8]. It was observed that the relative positioning of voids can affect the collapse temperatures and hotspot intensity of each of the voids in a cluster. There has been little previous work connecting meso-scale void-void interaction effects to macro-scale HE sensitivity. In early work, Nichols and Tarver [20] accounted for meso-structural information in a phenomenological reactive burn model base on a statistical distribution of hot spot sites To estimate the time to complete reaction for a spherical hotspot growing outward, their calculations assumed that the hotspot diameter increased to eight times its initial value [20]. By this measure, only small particles of HMX were found to react to completion. For large particles or hotspot spacings, the time to consume the intervening material during the growth phase can be larger than the reaction zone and therefore do not fully contribute to detonation. In fact, Nichols and Tarver argue that the critical hotspot classification, which relies on diffusion and reaction alone as potential mechanisms for hotspot growth, yields growth rates that are rather low for the reaction zones. They suggest [20] that there must be some other mechanism involved in the growth phase that can accelerate the reaction front growth rate. Tarver and Nichols [20] argued that other mechanisms such as grain fracture could be responsible for the high rates of reaction expected, where fracture provides increased surface area for hotspot growth. A similar argument was made by Saurel and coworkers [21] in their work on multiscale modeling of SDT in plastic bonded explosives (PBX). To match the experimental sensitivity response, as measured by pop-plots, Massoni et al. [22] enhanced the growth rate significantly by positing an interfacial amplification factor (IAF). The IAF depended exponentially on the pressure and led to rapid enhancement of the growth of reactions at higher pressures. Like Tarver and Nichols [20], increased fracture was assumed to be due to the high pressures and temperatures that arise during shock compression. While fracture at high pressures and temperatures as well as void-void interactions could be plausible mechanisms for enhancing the growth of reaction fronts, our current understanding of the effectiveness of such mechanisms in enhancing heat release rates in HEs remains incomplete. The present paper addresses the knowledge gap with regard to void-void interactions; damage and fragmentation of crystals is not considered in this work.

### 1.2 Focus and organization of the paper

The central questions that are addressed by this work are as follows:

1. The last two terms on the right hand side of the surrogate model in equations (1) and (2) above, i.e. $f_{ignition}^{shape}(AR,\theta)$ or $f_{growth}^{shape}(AR,\theta)$ and $f_{ignition}^{v-v}(\phi)$ or $f_{growth}^{v-v}(\phi)$, quantify effects of void shape effects and void-void interactions. Can surrogate models be constructed that quantify these modifying functions?
2. What shock physics or collapse mechanisms at the meso-scale lead to the observed topology of the resulting hypersurfaces $f^{shape}(AR,\theta)$ and $f^{v-v}(\phi)$?

This paper performs high-resolution meso-scale simulations of reactive void collapse in shocked HMX material to seek answers to these two questions.

The paper is organized as follows. Section 2 outlines the methods used for the computational analysis, where section 2.1 is a brief overview of the numerical methods and section 2.2 the methods used to construct the functions $f_{ignition}^{shape}(AR,\theta)$, $f_{growth}^{shape}(AR,\theta)$, $f_{ignition}^{v-v}(\phi)$, and $f_{growth}^{v-v}(\phi)$. Section 3 presents the results, which include the surrogate models for shape in Section 3.1 and surrogate model for void field in Section 3.2. Where Section 3.1.3, highlights physical explanations for the behavior of the surrogate models for shape i.e. $f_{ignition}^{shape}(AR,\theta)$, $f_{growth}^{shape}(AR,\theta)$. Similarly, the physics of the void-void interactions explaining the $f_{ignition}^{v-v}(\phi)$, and $f_{growth}^{v-v}(\phi)$ surrogates are elucidated in Sections 3.2.3 and 3.2.4. Conclusions and directions for future work are presented in Section 4.

# 2 METHODS

## 2.1 Numerical methods for solving governing equations and metamodel construction

A levelset-based sharp interface Cartesian grid Eulerian framework is used to perform void collapse simulations. The governing equations, along with material and reaction models, are presented in the Appendix (Section 7). The governing equations apply to shock-induced deformation of HMX with voids in the solid material treated as vacuum. The condensed phase is modeled as an elastic-perfectly plastic material to calculate deviatoric stresses, with a Mie-Gruneisen equation state for pressure. The Tarver Nichols 3-equation reactive model for HMX is used for calculating the chemical decomposition and combustion of the material. A detailed description of the overall framework is available from previous work [18, 19, 23-25].

The conservation laws of mass, momentum, and energy, along with the evolution of deviatoric stresses and species transport equations are spatially discretized using a 3rd order ENO scheme [26]. The time integration is performed using a 3rd order Runge-Kutta scheme. Narrow-band levelset tracking [27] is used in the current framework to sharply track the material interfaces. The levelset value denoted as $\varphi(x,t)$ at a point is the signed normal distance from the interface boundary. The value of levelset, $\varphi(x,t)$is less than zero within the void regions and greater than zero outside the void regions. The interface is implicitly determined by the contour $\varphi(x,t)=0$. The use of the levelset function allows handling of large deformation of interfaces that occurs during void collapse events. The interfacial conditions between the HMX and void are modeled by applying the free surface conditions at the interface using a modified ghost fluid method [28]. Detailed descriptions of the numerical algorithms, levelset implementation and interface treatment are provided in previous works [18, 19, 23-25].

Surrogate modeling is used to represent the hotspot ignition and growth rates, which are the quantities of interest (QoIs) that are supplied to macro-scale simulations in the ignition and growth framework. Surrogate modeling involves the estimation of an unknown function,$f(\boldsymbol{x})$, based on input data at a sparse set of discrete and distinct points $\mathbf{x}_j$ ($j$=1,…$N$). The set of the known values of the function and their locations, i.e. the set ($\mathbf{x}_j$), are the inputs for the surrogate model and are used to estimate the unknown function $f(\boldsymbol{x})$ in the parameter space. The point $\boldsymbol{x}_0$, where the value of $f(\boldsymbol{x})$ is of interest, is called the "probe point" of the surrogate model, while the value of the function at the probe point, $f(\boldsymbol{x}_0)$, is called the output of the surrogate model. The modified Bayesian Kriging (MBKG) approach for constructing the function $f(\boldsymbol{x})$ given ($\mathbf{x}_j$) was shown in previous work [29-31] to be more efficient than other methods, such as neural networks [29, 32] and polynomial stochastic collocation[33], for model representation using limited input data sets. Therefore, the MBKG approach is used to learn surrogate models from a small ensemble of meso-scale numerical experiments. Further details on the MBKG approach and machine-learning from meso-scale simulations can be found in [3] and [30].

## 2.2 Quantifying void shape and void-void interaction effects

In Part I [3], the functions $\dot{F}^{scv}_{ignition}$ and $\dot{F}^{scv}_{growth}$ were constructed using meso-scale simulations. As seen from Eqns. (1) and (2), $\dot{F}^{scv}_{ignition}$ and $\dot{F}^{scv}_{growth}$ serve as the foundation for modeling $\dot{F}_{ignition}$ and $\dot{F}_{growth}$. The functions $f^{shape}_{ignition}, f^{shape}_{growth}$ and $f^{v-v}_{ignition}, f^{v-v}_{growth}$ can be considered modifiers of the behavior of a single cylindrical void. $f^{shape}_{ignition}$ and $f^{shape}_{growth}$ account for the deviation of a representative void from cylindrical shape. $f^{v-v}_{ignition}$ and $f^{v-v}_{growth}$ account for the effect of a surrounding field of voids of void fraction $\phi$. As shown later in this paper, $f^{shape}_{ignition}$ or $f^{shape}_{growth}$ as well as $f^{v-v}_{ignition}$ or $f^{v-v}_{growth}$ assume values that depart significantly from 1, implying significant modifications of sensitivity due to elongated (non-cylindrical) void shapes or void-void interactions [8, 34].

Note that, in general, the effects of void shape and void-void interactions cannot be decoupled from the effects of void size and shock loading. Therefore, Eqns. (1) and (2) reflect limitations of the current analysis, resulting from the assumption that void shape and void interaction effects are both orthogonal to each other and to the effects of loading and void size. This assumption is made in the interest of tractability as explained in Part I [3, 35]. Therefore, the modifying functions in this work are only constructed for one loading condition, i.e. shock pressure of $P_s^o = 9.5\ GPa$ and $\tau_s^o = 3.77\ ns$ and one void size $D^0_{void} = 10\mu m$. The implications of the assumed form of Eqns. (1) and (2) for prediction of sensitivity of pressed HMX, when void shape, size, and spacing fully couple with loading, are being examined in ongoing work.

### 2.2.1 *Techniques for collecting data on void collapse-induced hotspot ignition and growth*

#### 2.2.1.1 *Defining a "representative" void*

As mentioned above, there are two distinct situations of interest in this work: 1) the ignition and growth of hotspots due to collapse of an isolated non-cylindrical void, and 2) the ignition and growth of hotspots due to collapse of a void embedded in a cluster of neighboring voids. The simulation setup and details of data collection for these two situations are quite different, as illustrated in the computational setups shown in Figure 2, Figure 3, and Figure 4. However, the analysis can be unified by adopting a framework common to the two scenarios, namely a "representative void" viewpoint. In this viewpoint, we calculate hotspot quantities of interest (QoIs) resulting from the collapse of the representative void. For a single void system, whether cylindrical or elongated, the definition of a "representative" void is straightforward. For a void-field, data is collected from individual control volumes enclosing the voids and then averaging over all the control volumes in the field to calculate the response of the "representative" void. The sensitivity of this representative void is then compared to that of an isolated single cylindrical void to quantify the effect of void shape and void-void interactions.

#### 2.2.1.2 *Defining control volumes for collecting data on hotpots*

In all cases, several hotspot QoIs comprise the set $\mathbb{Q}$. The procedure for extracting the set $\mathbb{Q}$ is described below for a single void and a field of voids. The control volume $cv$ for the single void case shown in Figure 2 and Figure 3 has length $l_{cv}$ on each side, the $l_{cv}$ values are provided in the respective figures. Similarly, for a void field, each individual control volume $cv$ in Figure 5 has a length $l_{cv}$ and contains approximately one void per control volume, such that:

$$l_{cv} = \sqrt{\frac{\pi\, D^2_{void}}{4\,\phi} * \frac{N_{voids}}{N_{cv}}} \tag{3}$$

where $D_{void}$ is the diameter of each of the $N_{voids}$ that make up the void volume fraction $\phi$ in the domain. Note that because the voids are arranged randomly in the domain it is likely that a control volume may not fully enclose a void at all times or in some situations may enclose more than one void. The averaging procedure described below for the field of voids takes account such contingencies. In the above, $N_{cv}$ is the total number of non-overlapping control volumes for a given volume fraction $\phi$. For simplicity of constructing $CV$, $N_{voids} = N_{cv}$ and $\sqrt{N_{cv}}$ is rounded to the nearest integer values. The length $L_{CV}$ of the larger control volume $CV$ enclosing the entire field of voids is therefore:

$$L_{CV} = l_{cv} \times \sqrt{N_{cv}} \tag{4}$$

Note that application of shock loading leads to the advection of the voids due to the particle velocity field following shock passage. To enclose all the voids throughout the computation, the larger control volume $CV$ (and thereby each $cv$) is translated at a speed of $V_{CV}$ in the direction of shock propagation. The speed $V_{CV}$ at time instant $t$ is the average x-component of the velocity of the material within the control volume $(u_x)$ and is given by:

$$V_{CV}(t) = \frac{\int_{\Omega_{CV}} \rho(t)\, u_x(t)\, d\Omega_{CV}}{\int_{\Omega_{CV}} \rho(t)\, d\Omega_{CV}} \tag{5}$$

where ρ is the local density. The individual control volumes $cv$ therefore move along with the voids as they collapse to form hotspots. During this process, $\mathbb{Q}$ are recorded for the individual control volumes $cv$ numbered $i_{cv}$ ( $1, \dots, N_{cv}$).. Other averaged field quantities are also calculated in similar fashion. For example, the total kinetic energy in a control volume is calculated as:

$$KE\,(i_{cv}; t) = \frac{1}{2} \int_{\Omega_{cv}} \rho(x,t)\, u(x,t)^2\, \Omega_{cv} \tag{6}$$

where $u(x,t)$ is the particle velocity in the x-direction at location $x$ and time $t$ within the domain. The time instant of entry of a shock wave into the control volume is recorded as $\tau_{shift}(i_{cv})$.

$$\tau_{shift}(i_{cv}) = min\,(t) \;\text{ such that }\; \Delta KE\,(i_{cv}; t) > 0 \tag{7}$$

where $\Delta KE\,(i_{cv}; t)$ is the instantaneous rise of the kinetic energy:

$$\Delta KE\,(i_{cv}; t) = KE\,(i_{cv}; t) - KE\,(i_{cv}; t - \Delta t) \tag{8}$$

The shifted time $t^*(i_{cv})$ for an individual control volume is defined as:

$$t^*(i_{cv}) = t - \tau_{shift}(i_{cv}) \tag{9}$$

Note that for the single control volume depicted in Figure 2 and Figure 3, $N_{cv} = 1$, and $V_{CV} = 0$, and $\tau_{shift} = 0$.

#### 2.2.1.3 Hotspot $QoIs$

The following set $\mathbb{Q}$ and averaged field quantities in the overall control volume $CV$ are relevant to the analysis presented in this work, and will be used to calculate the hotspot ignition and growth rates in Section 2.2.3.

1. Reaction product fraction, $F$: The dynamics of void collapse leads to hotspots where HMX decomposes into the gaseous reaction products. The product species mass fraction is denoted by $Y_4(x,t)$ and is evolved using the conservation equations for species (Appendix, Eq. (33)). The rate of chemical decomposition of HMX within $cv$ is calculated by accumulating the mass of the final gaseous products via:

$$M_{reacted}(i_{cv};t) = \int_{\Omega_{cv}} \rho(x,t)\, Y_4(x,t)\, d\Omega_{cv} \tag{10}$$

   $M_{reacted}$ is calculated for a control volume $cv$ of area $\Omega_{cv} = l_{cv}^2$. In line with Springer et al. [17] and Sen et al. [2], the reaction product fraction is the ratio,

$$F(i_{cv};t) = \frac{M_{reacted}(i_{cv};t)}{M_{void}} \tag{11}$$

   Here $M_{void}$ is the mass of HMX that would fill the void area given by $\frac{\pi}{4}{D_{void}}^2\rho_{HMX}$, where $\rho_{HMX} = 1900\ Kg/m^3$.

2. The rate of change of reaction product fraction, $\dot{F}$, is calculated as:

$$\dot{F}(i_{cv};t) = \frac{\Delta F(i_{cv};t)}{\Delta t} \tag{12}$$

   where $\Delta F$ is the change in the gaseous product fraction $F$ over the time interval $\Delta t$. In the MES-IG model [2] this reaction product formation rate is used in the ignition-and-growth framework to quantify the consumption of HMX to product gases at the macroscale.

3. Hotspot area, $A_{hs}$: A hotspot in this study is defined as the region with temperature greater than the bulk post-shock temperature. The hotspot area $A_{hs}$ in each control volume at time $t$ is calculated as:

$$A_{hs}(i_{cv};t) = \int_{\Omega_{cv}} \zeta(x,t) d\Omega_{cv} \tag{13}$$

   $\zeta(x,t)$ is an indicator function that labels hotspot regions and is defined as:

$$\zeta(x,t) = \begin{cases} 1, & T(x,t) > T_{bulk} \\ 0, & T(x,t) \le T_{bulk} \end{cases} \tag{14}$$

$T(x,t)$ is the temperature in a computational cell at time $t$ and $T_{bulk}$ is the bulk temperature of the domain after shock passage. The hotspot area fraction $\hat{A}_{hs}$ is the hotspot area $A_{hs}$ per unit control volume area $\Omega_{cv}$.

$$\hat{A}_{hs}(i_{cv};t) = \frac{1}{\Omega_{cv}} \int_{\Omega_{cv}} \zeta(x,t) d\Omega_{cv} \tag{15}$$

$\hat{A}_{hs}(i_{cv};t)$ reaches a maximum value of 1 when all the material within the control volume $cv$ is above the bulk temperature.

4. The hotspot temperature, $T_{hs}$, is the average temperature of the hotspot in the control volume:

$$T_{hs}(i_{cv};t) = \frac{\int_{\Omega_{cv}} \rho(x,t)\,\zeta(x,t)\,T(x,t) d\Omega_{cv}}{\int_{\Omega_{cv}} \rho(x,t)\,\zeta(x,t) d\Omega_{cv}} \tag{16}$$

5. The void area, $A_{void}$, is obtained from:

$$A_{void}(i_{cv};t) = \int_{\Omega_{cv}} \xi(x,t)\, d\Omega_{cv} \tag{17}$$

where $\xi(x,t)$ is the indicator function distinguishing void regions using the levelset field:

$$\xi(x,t) = \begin{cases} 1, & \varphi(x,t) < 0 \\ 0, & \varphi(x,t) > 0 \end{cases} \tag{18}$$

and $\varphi(x,t)$ is the levelset field value as defined in the Section 2.1.

The hotspot QoIs in Eqs. (10-19) are used to obtain the ignition and growth rates and analyze the behavior of hotspots. $\mathbb{Q}(i_{cv};t^*)$ are then averaged over all the control volumes, as in Eq. (19) below. This averaged QoI then applies to a representative void in the field of voids shown in Figure 3. For the case of a field of voids, spatially averaged quantities $\mathbb{Q}$ are obtained as follows:

$$\mathbb{Q}\,(t^*) = \frac{1}{N_{cv}} \sum_{i_{cv}=1}^{N_{cv}} \mathbb{Q}(i_{cv};t^*) \tag{19}$$

Note the use of the shifted time $t^*$ discussed previously, which has significance for the random void fields.

#### 2.2.1.4 Calculating the hotspot ignition and growth rates $\dot{F}_{ignition}$ and $\dot{F}_{growth}$

Similar to the calculation for a single cylindrical void [3, 17], the ignition rate $\dot{F}_{ignition|i_{cv}}$ for control volume $i_{cv}$ is obtained as the time rate of change of $F$ between points 1 and 2. Therefore:

$$\dot{F}_{ignition|i_{cv}} = \left(\frac{F_2 - F_1}{t_2^* - t_1^*}\right)_{i_{cv}} \tag{20}$$

For the void field, the averaged ignition rate is calculated from:

$$\dot{F}_{ignition} = \frac{1}{N_{cv}} \sum_{i_{cv}=1}^{N_{cv}} \dot{F}_{ignition|i_{cv}} \tag{21}$$

The hotspot growth rate is calculated similarly as the time rate of change between points 3 and 4. Thus $\dot{F}_{growth|i_{cv}}$ and $\dot{F}_{growth}$ are calculated as

$$\dot{F}_{growth|i_{cv}} = \left(\frac{F_4 - F_3}{t_4^* - t_3^*}\right)_{i_{cv}} \tag{22}$$

$$\dot{F}_{growth} = \frac{1}{N_{cv}} \sum_{i_{cv}=1}^{N_{cv}} \dot{F}_{growth|i_{cv}} \tag{23}$$

The above procedure to obtain $\dot{F}_{ignition}$ and $\dot{F}_{growth}$ is common to both the elongated void (for which $N_{cv}$=1) and for the void field where the averaging is performed over all the control volumes $cv$. For the elongated voids, once the hotspot ignition and growth rates are found from Eq. (24) and Eq. (26) for various values of $AR$ and $\theta$, the modifying functions $f_{ignition}^{shape}(AR,\theta)$ and $f_{growth}^{shape}(AR,\theta)$ are obtained by normalizing the $\dot{F}_{ignition}$ and $\dot{F}_{growth}$ by the corresponding rates for the single cylindrical void:

$$f_{ignition}^{shape}(AR,\theta) = \frac{\dot{F}_{ignition}(P_s^o, \tau_s^o, D_{void}^o, AR, \theta)}{\dot{F}_{ignition}^{scv}(P_s^o, \tau_s^o, D_{void}^o)} \tag{24}$$

$$f_{growth}^{shape}(AR,\theta) = \frac{\dot{F}_{growth}(P_s^o, \tau_s^o, D_{void}^o, AR, \theta)}{\dot{F}_{growth}^{scv}(P_s^o, \tau_s^o, D_{void}^o)} \tag{25}$$

$\dot{F}_{ignition}^{scv}$ and $\dot{F}_{growth}^{scv}$ for the single cylindrical void are calculated as described in Part I [3]. For a field of voids, $\dot{F}_{ignition}$ and the $\dot{F}_{growth}$ must be averaged for the representative void for various $\phi$ values, and then the corresponding modifying functions $f_{ignition}(\phi)$ and $f_{growth}(\phi)$ can be calculated:

$$f_{ignition}^{v-v}(\phi) = \frac{\dot{F}_{ignition}(P_s^o, \tau_s^o, D_{void}^o, \phi)}{\dot{F}_{ignition}^{scv}(P_s^0, \tau_s^0, D_{void}^0)} \tag{26}$$

$$f_{growth}^{v-v}(\phi) = \frac{\dot{F}_{growth}(P_s^o, \tau_s^o, D_{void}^o, \phi)}{\dot{F}_{growth}^{scv}(P_s^0, \tau_s^0, D_{void}^0)} \tag{27}$$

As an example, Figure 6 shows simulation results for void area, hotspot temperature, hotspot area, and product mass fraction for an elongated void of $AR = 22$ and $\theta = 15°$ at the reference loading conditions $P_s^0$ and $\tau_s^0$. Ignition, bounded by points 1 and 2, is defined by the void collapse process shown in Figure 6 (a), such that point 1 is the time at which the void area is 90% of its original value and point 2 is the time at which the void has collapsed and the void area is zero [3]. Growth, bounded by points 3 and 4, is defined by the hot spot growth process shown in Figure 6 (c), such that points 3 and 4 and the times at which the hot spot area is 1.8 and 2.0 times larger than the area of the hotspot at the end of the ignition phase, respectively [3, 17].

Similarly, Figure 8 shows simulation results for void area, hotspot temperature, hotspot area, and product mass fraction for a field of voids of $\phi = 5$ % at the reference loading conditions $P_s^0$ and $\tau_s^0$. Ignition is defined by the void collapse process described for the elongated void case above [3]. Growth for the field of voids is also defined by the hot spot growth process used for the elongated void, but for the void field, points 3 and 4 are defined as the times at which the hot spot area is 1.3 and 1.5 times larger than the area of the hotspot at the end of the ignition phase, respectively. The growth rates are calculated at these points of time because hotspot of these sizes are contained well within the control volumes before the hotspots start to merge with each other.

## 3 RESULTS

The above methods are used below to quantify void shape and void-void interaction effects. In the following section, the surrogate model construction is first presented, followed by an analysis of the underlying physics of hotspot dynamics for non-cylindrical shaped voids and multiple void collapse in a porous HMX matrix.

### 3.1 Construction of the $f_{ignition}^{shape}(AR, \theta)$ and $f_{growth}^{shape}(AR, \theta)$ surrogate model

To obtain $f_{ignition}^{shape}(AR, \theta)$ and $f_{growth}^{shape}(AR, \theta)$, reactive void collapse simulations are performed for elongated voids in the computational setup shown in Figure 3. An elongated void of length $l_{void}$ and thickness $b$, oriented at an angle of $\theta$ with respect to the x-axis, is embedded in an otherwise uniform HMX matrix of size $100\ \mu m \times 50\ \mu m$. The aspect ratio of the void is defined as $AR = l_{void}/b$. The edges of the void have a fillet radius of $0.5b$. The void area of the elongated voids corresponds to that of a cylindrical void of diameter $D_{void}^0$. As discussed previously, the functions $f_{ignition}^{shape}(AR, \theta)$ and $f_{growth}^{shape}(AR, \theta)$ are calculated for the specific loading condition $P_s^0$ and $\tau_s^0$.

#### *3.1.1 Data collection for void shape variations*

Reactive meso-scale simulations are performed to obtain $\dot{F}_{ignition}$ and $\dot{F}_{growth}$ for $AR = \{1, 4, 7, 10, 13, 16, 19, 22\}$ and $\theta = \{0°, 15°, 30°, 45°, 60°, 75°, 90\}$. Figure 6 shows results for void area, hotspot temperature, hotspot area, and product mass fraction for an elongated void of $AR = 22$ and $\theta = 15°$. The values of the modifier functions $f_{ignition}^{shape}$ and $f_{growth}^{shape}$ obtained using Eqs. (24) and (25) are used as training data for the MBKG algorithm [31] to construct the surrogate models. As shown in previous work [30], reliable surrogate hypersurfaces are produced by the MBKG algorithm using roughly eight parameter values along each parameter dimension. The surrogate models for $f_{ignition}^{shape}$ and $f_{growth}^{shape}$ therefore use 56 high fidelity reactive void collapse simulations.

#### *3.1.2 Surrogate Models: $f_{ignition}^{shape}(AR, \theta)$ and $f_{growth}^{shape}(AR, \theta)$*

Figure 7 shows the modifier functions $f_{ignition}^{shape}(AR, \theta)$ and $f_{growth}^{shape}(AR, \theta)$ with 95% credible sets obtained from the MBKG fit. For slender voids with $AR > 4$ and orientations $0° < \theta < 45°$, the hotspot ignition rate is amplified above that seen in a corresponding cylindrical void of the same void area, as reported in Figure 7 (a) and Table 1. However, for $\theta > 45°$ the modifying shape function decreases to less than one. In fact, for voids with $AR > 19$ and $\theta > 75°$, the value of $f_{ignition}^{shape}(AR, \theta) \approx 0$. Therefore, an elongated void at this unfavorable orientation with respect to the shock will fail to ignite, even at conditions within the critical regime for the corresponding single cylindrical void.

The $f_{growth}^{shape}(AR,\theta)$ in Figure 7 (b) and reported in Table 2 is also amplified to a modest extent for the elongated voids lying in the range of $AR > 4$ and $\theta < 45°$, where the maximum value is reported for $f_{growth}^{shape}(AR = 22, \theta = 15°) = 1.75$. Note that void shape more strongly effects the likelihood of ignition of the hotspot than the subsequent growth phase $f_{growth}^{shape}(AR,\theta)$. The physics underlying the enhancement or inhibition of the ignition and growth rates for slender voids is explained next.

### *3.1.3 Physics of ignition of hot spots due to void shape effects*

From Figure 7, the ignition and growth rates of elongated voids are observed to be higher than the corresponding cylindrical voids for $0° < \theta < 45°$ and $4 < AR < 22$ [34]. To understand the physics in this range of $\theta$ and $AR$, the mechanism of collapse and hotspot characteristics such as shape, size, and temperature distributions for an elongated void ($AR = 22, \theta = 15°$), are compared with a cylindrical void ($D_{void}^0$) for the reference loading parameters $P_s^0$ and $\tau_s^0$.

Figure 9 shows the contour plots of vorticity, temperature, and final gaseous product mass fraction for both voids after collapse. During the collapse of the elongated void the progressive closure of the void by repeated pinching along its length leads to the formation, propagation, and self-strengthening of blast waves [7]. Therefore, collapse of the elongated voids progresses with increasing collapse temperature along the length of the void under the influence of self-strengthened blast waves. The repeated pinching also leads to the generation of high gradients of pressure and density across the void thickness. These gradients give rise to the formation of baroclinic vortices, as can be seen from vorticity contour plots of Figure 9 (a). Comparison of Figure 9 (a) and (c) show high temperature regions at the core of the baroclinic vortices. This is also observed for cylindrical voids from Figure 9 (b) and (d), where the high temperature regions coincide with the high concentration of vorticity at the location of secondary lobe collapse [36]. For both the voids, in the high vorticity regions, mixing due to enhanced contact area of hot reacted material and the surrounding colder unreacted HMX lead to increased reaction rates and reaction completion. This can be seen from Figure 9 (e) and (f), where the contour plots of final reaction product show that the vortex cores are collocated with maximum reaction completion. The vorticity concentration for the elongated void is higher and mixing occurs all along the void surface, i.e. over a large surface area compared to the cylindrical void. Therefore, collapse of elongated voids with self-strengthened blast waves, high vorticity concentration, and mixing lead to enhanced hotspot ignition rates compared to cylindrical voids. The difference in the dynamics of void collapse behavior of elongated and cylindrical voids can lead to an order of magnitude amplification of hotspot ignition rate, which is reflected in the $f_{ignition}^{shape}(AR,\theta)$ modifier function as can be seen from Figure 7 (a) and (c).

Once the collapse of voids has occurred, the rate of reaction front growth is dependent on the shape, size, and temperature of the hotspot. An estimate of the hotspot morphology can be obtained from the temperature contours for both the cylindrical and elongated voids as shown in Figure 9 (a) and (b). The hotspot due to an elongated void reflects its original shape, leading to an enhanced burning front surface as can be seen from Figure 9 (b). For the cylindrical void, the hotspot is mushroom-shaped with the high temperature region concentrated at the collapse locations of the two symmetrical lobes. The overall hotspot for the cylindrical void is more compact than that of the elongated void. Therefore, the larger burn surface area of the elongated hotspot relative to the more compact hotspot formed by the cylindrical void is expected to lead to a higher $\dot{F}_{growth}$ for elongated voids. The relationship between the reaction growth rate, $\dot{F}_{growth}$ and surface area of the hotspots is analyzed further by comparing the variation of $f_{growth}^{shape}(AR,\theta)$ and hotspot surface area of elongated voids with varying $AR$. The variation of $f_{growth}^{shape}(AR,\theta)$ and the ratio of hotspot surface area for the elongated and cylindrical void are shown in Figure 10. Both $f_{growth}^{shape}(AR,\theta)$ and burn surface area ratio increase with $AR$ showing a direct relationship between the reaction growth rates and the surface area of the voids. However, the $f_{growth}^{shape}(AR,\theta)$ does not increase as strongly with $AR$ as the hotspot surface area. Thus, the surface area ratio for the elongated void with $AR = 22$ is 2.76 while the $f_{growth}^{shape}(AR = 22, \theta = 15°) = 1.75$. This is due to the vortical rollup shown in Figure 9 (d), where the hotspot formed by the elongated void collapse has a larger burn front surface area than would be expected if the hotspot shape maintained the original cylindrical shape.

In summary, void shapes that deviate from the idealized cylindrical shape can enhance or suppress the sensitivity of a porous HE, as indicated by the factors $f_{ignition}^{shape}(AR,\theta)$ and $f_{growth}^{shape}(AR,\theta)$in Figure 7. The void shape has a stronger effect on the hotspot ignition rate than on the growth rate.

## 3.2 Construction of the $f_{ignition}^{v-v}(\phi)$ and $f_{growth}^{v-v}(\phi)$ surrogate model

The effect of void-void interactions is quantified using the modifier functions $f_{ignition}^{v-v}(\phi)$ and $f_{growth}^{v-v}(\phi)$, calculated using Eqns. (26) and (27). Figure 3 shows the setup for simulating the effects of void-void interactions. A field of cylindrical voids is subject to shock loading from the left face of the domain. The numerical simulations are performed at the reference loading conditions, $P_s^o$and $\tau_s^o$. The domain size is constant at $400\mu m$ x $300\mu m$ with voids of diameter $D_{void}^o$. The volume fraction $\phi$ is varied by changing the number of voids in the domain. For each $\phi$, an average inter-void distance is maintained. The *rand* function in MATLAB [37] is used to place voids of specified diameter in the control volume *CV*. An additional criterion for the void placement is the minimum inter-void distance: no two voids are allowed to be closer than a specified minimum inter-void distance. An iterative procedure is used to obtain various arrangements until the highest possible value of the minimum inter-void distance is achieved for a specified $\phi$. Arranging the voids through this criterion ensures the voids are distributed randomly but more or less homogenously throughout the sample. The effect of local variations in $\phi$, i.e. void clustering effects, are left for future study.

### *3.2.1 Data collection and averaging for a field of voids*

The process of collecting data from the moving control volume and calculating averaged quantities is illustrated in Figure 4 for a void field with $\phi = 5\%$. The large control volume $CV$ enclosing all the voids is sub-divided into $N_{cv} = 25$ smaller control volumes with numbers $i_{cv}$ indicated in the top left corner of each control volume $cv$.

Figure 11 (a) shows the variation of total kinetic energy $KE$ with time (Eqn. (7)) for the centerline control volumes numbered $i_{cv}$= 11 to 15. Since the shock is applied at the left side of the domain, control volumes located further from the left boundary have higher values of $\tau_{shift}$. The peak in $KE$ is highest for $i_{cv} = 11$ as it faces the imposed shock without any attenuation. The presence of voids and the traversal of the shock through the HMX matrix leads to attenuation of the shock wave for this value of $\phi$, thus the control volumes located further from the left boundary show lower peaks in $KE$.

Figure 11 (b) shows the variation of hotspot temperature $T_{hs}$ with time (Eqn. (16)) for the centerline control volumes. The peak in the hotspot temperature corresponds to the instant of void collapse as discussed in Part 1 [3]. The control volumes further away from the left face of the domain experience an attenuated shock for this value of $\phi$, and therefore show smaller peak values of $T_{hs}$.

Figure 11 (c) shows the variation in time of the averaged product mass fraction $F$ (Eqn. (11)), which is a measure of the progress of chemical reactions for a particular control volume. The values of $F$ for control volumes 12 and 13 are larger than control volumes 14 and 15 because control volumes 14 and 15 are farther from the left face. Though control volume 11 is closest to the left side of the domain, $F$ and $T_{hs}$ are low because the control volume contains only a partial void. Figure 11 (d) shows the variation of $F$ for the centerline control volumes as a function of the shifted time, which shows the curves for the different control volumes overlap. The black line in Figure 11 (d) shows the variation of the averaged quantity $F(t^*)$ with time, which is obtained using Eqn. (19) by averaging the data over all the control volumes ($N_{cv} = 25$ for $\phi = 5\%$).

### 3.2.2 *Surrogate Models:* $f_{ignition}^{v-v}(\phi)$ *and* $f_{growth}^{v-v}(\phi)$

Figure 12 (a) and (b) show the $\dot{F}_{ignition}$ and $\dot{F}_{growth}$ in the field of voids, obtained over the range of void fractions $\phi = \{2\%, 5\%, 7\%, 10\%, 12\%, 15\%, 20\%\}$. The void fraction $\phi$ has a significant effect on the ignition and growth rates. In general, void-void interaction effects lead to higher values of hotspot ignition and growth rates when compared to an isolated single cylindrical void, shown by the modifying functions $f_{ignition}^{v-v}(\phi)$ and $f_{growth}^{v-v}(\phi)$ in Figure 12 (c) and (d). The $f_{ignition}^{v-v}(\phi = 2\%) \approx 1$, indicating that the $\dot{F}_{ignition}$ is similar to that of a single isolated void. On the other hand, the $f_{growth}^{v-v}(\phi = 2\%) = 0.6$, indicating slower hotspot growth rates compared to that of a single void. At $\phi = 2\%$, the attenuation of the lead shock wave as it passes through the void field leads to the formation of weaker hotspots at downstream voids. Furthermore, for $\phi = 2\%$ the voids are far enough apart that the blast waves produced as neighboring voids collapse lose their intensity before they reach the surrounding voids. The $f_{ignition}^{v-v}(\phi = 20\%) = 22.5$ and $f_{growth}^{v-v}(\phi = 20\%) = 2.25$, indicating an increase in blast wave interactions at higher values of $\phi$, i.e. as the void density increases. In the following section the physics of void-void interaction effects on the hotspot formation and growth mechanisms is discussed.

### 3.2.3 *Physics of void-void interactions: effect of blast waves from neighboring void collapses on hotspot growth rates*

To elucidate the effects of void volume fraction on the blast wave interactions, a comparison is made between the lowest volume fraction of $\phi = 2\%$ and the highest volume fraction of $\phi = 20\%$ for the reference loading condition. Figure 13 shows the pressure and temperature contour plots for a sample with $\phi = 2\%$ at various time instants. Figure 13 (a) shows the pressure at early time where the shock has propagated through the first few void fractions with no attenuation. The blast wave interacts with the lead shock front to form a Mach stem [12]. Figure 13 (b), shows that the temperature for the central void reaches 3000K, which is above the bulk temperature of 500K in the shocked HMX. The blast waves from the void collapse do not have a significant effect on neighboring voids because the inter-void spacing is large ($\sim 45\ \mu m$). The intensity of the circular blast waves produced by void collapse weakens rapidly as the waves spread out from the point of collapse. The large void spacing also allows for rarefactions to counteract compression, shown in Figure 13 (c) as the shock is nearly through the void bed. For the same time, Figure 13 (d) shows that the hotspots formed in the later (downstream) void collapse events are at temperatures close to 1500K. At the latest time considered, Figure 13 (e) shows the bulk pressure after the shock passes the entire domain ranges from 3 to 4 GPa. The lead shock is attenuated as it propagates through the porous material, which lowers the blast wave intensity propagating radially from the points of collapse or by interactions with rarefactions. Therefore, the downstream voids result in weaker hotspots.

For $\phi = 20\%$ the physics of interaction of the reactive blast waves with the closely spaced voids is quite different, as seen in Figure 14. The pressure field at early time matches the loading condition, shown in Figure 14 (a). The first several columns of voids collapse to produce strong blast waves, which lead to pressures approaching 20 GPa, shown in Figure 14 (a). The temperatures produced in the hotspots due to the collapse of the first few columns of voids are in the range of 4000K, shown in Figure 14 (b). Since the inter-void spacing is small ($\simeq 5\ \mu m$), the high intensity reactive blast waves impact the nearby voids, collapsing them at higher pressures, producing even stronger blast waves. As shown in Figure 14 (c), this progressive reinforcement of blast waves causes the pressure in the center of the control volume to rise to about 26 GPa. Since the voids in the downstream regions collapse under higher pressures, intense hotspots form with temperatures in the range of 5000K, as shown in Figure 14 (d).

The hotspot temperatures reached for $\phi = 20\%$ are significantly higher than for $\phi = 2\%$. The higher temperatures result in rapid chemical energy release, which feeds energy into the blast waves thereby increasing the pressure even further. This process occurs rapidly, in a matter of 60 ns, and leads to the formation of a self-sustaining detonation wave front seen in Figure 14 (e) and (f). The pressure at the leading edge of the detonation front is ~32 GPa and the temperature is ~ 6000 K. Therefore, reactive blast waves produced through void collapse have a large impact on the ignition and growth rates at higher void fractions. Note that the highest void fractions simulated (for example, 20%) are unlikely to be abundant in typical as-produced HEs. However, high local volume fractions may be present in isolated regions in stochastic meso-structures conferring increased sensitivity at such locations. Furthermore, as conjectured by Nichols and Tarver [20] and Massoni et al. [22], they may arise due to fragmentation of crystals at high pressures, leading to significant acceleration of hotspot growth rates.

#### *3.2.4 Distinguishing the roles of shock focusing and chemical energy release in void-void interactions*

As discussed above, the intensification of the incident shock for high void fractions is caused by the energy supplied by the reactive blast waves that emanate from void collapse events. Therefore, both the reinforcement due to the increased pressures from the blast waves as well as the addition of chemical energy at hotspot sites contribute to the enhancement of ignition and growth rates in void fields. To distinguish the effect of chemical heat release from the energy localization due to blast waves alone, an inert case is simulated by switching off chemistry. The inert simulations will indicate whether shock focusing due to void collapse alone can lead to the observed intensification of hotspots.

Figure 15 shows an inert simulation for the reference loading condition. The voids in the upstream column have just collapsed in Figure 15 (a) and a series of blast waves is formed. The blast wave pressures are in the range of 16 GPa, higher than the 22 GPa observed for the reactive simulation. The temperature contours for the reactive and inert cases are also markedly different. Figure 14 (b) shows multiple hotspots with temperatures reaching 4000K, whereas the inert simulation results in Figure 15 (b) show smaller hotspots with temperature near 1600K. As the shock reaches the end of the void field in Figure 15 (c), most of the voids have collapsed but the attenuated shock front has a concave structure, where the maximum pressure is only about 7 GPa, in stark contrast with the maximum pressures of 24 GPa seen in Figure 14 (c) for the reactive simulation. Figure 15 (d) shows that the low temperature, smaller sized hotspots formed in the inert simulations have begun to diffuse. At the final time shown in Figure 15 (e) and (f), the lead shock wave is significantly attenuated and has a strength of around 5 GPa with hotspot temperatures in the 1200K range, both of which are significantly lower than the reported pressure and temperature for the same time in the reactive simulation. Therefore, it is clear the response in the reactive simulation is an effect of the coupling between the chemistry, void collapse induced shock focusing, and the incident shock wave, with chemical heat release playing a predominant role.

Figure 16 (a) and (b) show the *x-t* plots (colored with pressure levels) for the reactive and inert simulations, respectively, for $\phi = 20\%$. Recall the void field starts at $X = 50\mu m$ and extends to $X = 250\mu m$, as shown in Figure 5; the 1-D pressure data is extracted along the x-direction for $Y = 150\ \mu m$. In both the reactive and inert simulations, the reference loading pressure is shown, as expected, through $X = 50\mu m$. Once the shock enters the void field, the reactive simulation shows higher pressures than the inert simulation. At the end of the void field at $X = 250\mu m$, the reactive simulation shows a peak pressure of 56 GPa, whereas the inert simulation shows the maximum pressure at 3.5 GPa.

Figure 17 shows the comparison of the slope of the shock locus in the *x-t* plane for the inert and reactive simulations. For comparison, the shock locus for an inert shock traveling through bulk (i.e. void-free) HMX is also plotted. Before the shock enters the void field at $X = 50\mu m$, all three shock loci overlap. Within the void field, the bulk HMX shows the highest shock locus slope; thus the unhindered shock travels fastest. Upon exiting the void field, the shock travels fastest in the reactive simulation. The speed of the detonation wave in the reactive simulation is calculated from the slope of the shock locus as ≃10,100 m/s. This is approximately twice the speed of the shock front in bulk HMX, ≃5500m/s. Thus, the strengthening of void-void interactions in a high porosity void field is primarily due to the addition of chemical energy from hotspots. The localization of energy and shock focusing due to inert void collapse itself does not lead to the strengthening of the incident shock.

## 4 CONCLUSIONS

Surrogate models for void shape and void-void interaction effects are constructed using machine learning; these models can be used to close the macroscale models in a multi-scale simulation of pressed HMX [2]. The surrogate models are trained using the ignition and growth rates calculated from meso-scale simulations of reactive void collapse. The surrogate models $f_{ignition}^{shape}(AR,\theta), f_{growth}^{shape}(AR,\theta)$ and $f_{ignition}^{v-v}(\phi), f_{growth}^{v-v}(\phi)$ developed using the MBKG method act as modifier functions to the $\dot{F}_{ignition}^{scv}$ and $\dot{F}_{growth}^{scv}$ surrogate models developed from single void collapse simulations in Part I [3].

Elongated voids with deviations from cylindrical shape defined by the aspect ratio $AR$, and orientation $\theta$ are used to capture void shape effects. Elongated voids with $4 < AR < 22$ and orientations between $0° < \theta < 45°$ show amplified ignition and growth rates compared to the corresponding cylindrical void of the same void area, where the ignition rate shows an order of magnitude increase and the growth rate shows a modest, but non-negligible, increase. Two key reasons are identified for the enhanced ignition rates for hotspots due to elongated void collapse: a) repeated pinching of the void surface along the length of the elongated void leads to the collapse of the void under the influence of self-strengthened blast wave, b) the formation of strong baroclinic vorticity during the collapse of the elongated void leads to higher temperature rise and reaction completion. In the growth phase, the increase in the reaction growth rate for elongated voids is related to the larger hot spot burn surface area of the elongated voids. Since the difference in the burn front surface area of the hotspots for the elongated and cylindrical voids does not differ by an order of magnitude, there is only a modest increase in the $f_{growth}^{shape}(AR,\theta)$ function compared to the $f_{ignition}^{shape}(AR,\theta)$.

Spatially and temporally averaged QoIs are obtained to quantify the effects of void-void interaction on a representative void embedded in a field of voids. The averaging is performed in a moving control volume that encompasses all voids. The calculated QoIs are used to obtain the $f_{ignition}^{v-v}(\phi)$ and $f_{growth}^{v-v}(\phi)$ modifying functions. It is observed that the void volume fraction $\phi$ has a significant effect on the ignition and growth rates, where the blast wave interactions between neighboring voids at higher values of $\phi$ lead to higher temperature hotspots. At the highest void volume fraction considered, the ignition rate shows an order of magnitude increase and the growth rate shows a modest, but non-negligible, increase. The amplification in the ignition and growth for $\phi = 20\%$ leads to the formation of a meso-scale detonation front. While high void fractions (for example, 20%) are unlikely to be abundant in typical as-produced HEs, high local volume fractions may be present in stochastic meso-structures conferring increased sensitivity at such locations. Also, as conjectured by Nichols and Tarver [20] and Massoni et al. [22] high volume fractions may arise because of crystal fragmentation at high pressures, that lead to accelerated hotspot growth rates.

This work shows that in simulating the response of heterogeneous energetic materials at the macro-scale, a meso-informed closure model for energy localization should take into account void-void interactions and void shape effects. Both these features can alter the sensitivity of porous HEs significantly. Knowledge of the meso-structure of a HE can be obtained by imaging the meso-structure and deriving statistical information on the meso-morphology. While detailed meso-scale simulations of the response of imaged HE samples are now feasible [18], for macro-scale or process-scale sample sizes homogenized closure models will continue to be the norm in the foreseeable future. Such models can combine quantification of imaged meso-scale morphological metrics (such as void size, void fraction, void shape features etc.) and ignition and growth rate dependency on such metrics (via surrogate models developed in this work) to improve predictions of HE response.

## 5 ACKNOWLEDGEMENTS

The authors gratefully acknowledge the financial support from the Air Force Research Laboratory Munitions Directorate, Eglin AFB, under contract number FA8651-16-1-0005. The authors are also thankful to Dr. K.K. Choi at the University of Iowa and Dr. Nicholas J. Gaul at RAMDO LLC, Iowa City, for providing the computational code for the Modified Bayesian Kriging Method.

# 7 APPENDIX

## 7.1 Governing equations

The hyperbolic conservation laws for mass, momentum and energy are solved:

$$\frac{\partial \rho}{\partial t} + \frac{\partial (\rho u_i)}{\partial x_i} = 0 \tag{28}$$

$$\frac{\partial (\rho u_i)}{\partial t} + \frac{\partial \left(\rho u_i u_j - \sigma_{ij}\right)}{\partial x_j} = 0 \tag{29}$$

and

$$\frac{\partial(\rho E)}{\partial t}+\frac{\partial\left(\rho E u_j-\sigma_{ij}u_i\right)}{\partial x_j}=0 \tag{30}$$

where $\sigma_{ij}$ is the Cauchy stress tensor and is decomposed into volumetric and deviatoric component, i.e.:

$$\sigma_{ij}=S_{ij}-p\delta_{ij} \tag{31}$$

The deviatoric stress tensor, $S_{ij}$, is evolved using the following evolution equation.

$$\frac{\partial\left(\rho S_{ij}\right)}{\partial t}+\frac{\partial\left(\rho S_{ij}u_k\right)}{\partial x_k}+\frac{2}{3}\rho G D_{kk}\delta_{ij}-2\rho G D_{ij}=0 \tag{32}$$

where, $D_{ij}$ is the strain rate tensor, and $G$ is the shear modulus of material. First the deviatoric stresses are evolved using an elastic response and then mapped back to the yield surface using a radial return algorithm [38]. The yield surface is given by the function $f=S_e-\sigma_y$, where $S_e=\sqrt{\frac{3}{2}\left(S_{ij}S_{ij}\right)}$. The yield strength, $\sigma_y$ is taken to be a constant and set to $260\ MPa$ [39] for HMX, i.e. hardening, and visco-plastic effects are neglected in the meso-scale computational models. A detailed description of the governing equations and radial return algorithm is provided in previous work [18, 19, 23-25].

The chemical species are evolved in time by solving the conservation equation:

$$\frac{\partial\rho Y_i}{\partial t}+div\left(\rho\vec{V}\ Y_i\right)=\dot{Y}_i \tag{33}$$

where $Y_i$ is the mass fraction of the $i^{th}$ species and $\dot{Y}_i$ is the production rate source term for the $i^{th}$ species. The numerical stiffness in solving the reactive set of equations is circumvented by using a Strang operator splitting approach [40], where first the advection of species is performed using the flow time step to obtain predicted species values:

$$\frac{\partial\rho Y_i^*}{\partial t}+div\left(\rho\vec{V}^n\ Y_i^*\right)=0 \tag{34}$$

In a second step, the evolution of the species mass fraction due to chemical reactions is calculated:

$$\frac{dY_i^{n+1}}{dt}=\dot{Y}_i^{*n} \tag{35}$$

The species evolution Eqn. (35) is advanced in time using a 5$^{th}$-order Runge-Kutta Fehlberg method [41], which uses an internal adaptive time-stepping scheme to deal with the stiffness of the chemical kinetic equations.

### 7.2 Constitutive and Reaction Models

The pressure in the meso-scale in Eqn. (31) is obtained from a Birch-Murnaghan equation of state [39, 42], which can be written in the general Mie-Gruneisen form as:

$$p(\rho,e)=p_k(\rho)+\rho\Gamma_s[e-e_k(\rho)] \tag{36}$$

where

$$p_k(\rho) = \frac{3}{2} K_{T0} \left[ \left(\frac{\rho}{\rho_0}\right)^{-7/3} - \left(\frac{\rho}{\rho_0}\right)^{-5/3} \right] \left[ 1 + \frac{3}{4}(K'_{T0} - 4) \left[ \left(\frac{\rho}{\rho_0}\right)^{-2/3} \right] - 1 \right] \quad (37)$$

Void collapse under shock loading can lead to the melting of HMX; therefore thermal softening of HMX is modeled using the Kraut-Kennedy relation, $T_m = T_{m0}\left(1 + a\frac{\Delta V}{V_0}\right)$, with model parameters provided in the work of Menikoff et al. [39]. Once the temperature exceeds the melting point of HMX the deviatoric strength terms are set to zero. Furthermore, the specific heat of HMX is known to change significantly with temperature. The variation of specific heat is modeled as a function of temperature as suggested in [39].

The chemical decomposition of HMX is modeled using a 3-step mechanism proposed by Tarver et al. [43]. A detailed description of the implementation of the 3-step model in the current numerical framework is presented in previous work [34]. Here, a brief overview of the reaction model and its implementation is provided.

Chemical decomposition of HMX takes place in three steps involving four different species:

Reaction 1:

$$HMX\ (C_4H_8N_8O_8) \rightarrow \text{fragments}\ (CH_2NNO_2) \quad (38)$$

Reaction 2:

$$\text{fragments}\ (CH_2NNO_2) \rightarrow \text{intermediate gases}\ (CH_2O, N_2O, HCN, HNO_2) \quad (39)$$

and

Reaction 3:

$$2 \times \text{intermediate gases}\ (CH_2O, N_2O, HCN, HNO_2) \rightarrow \text{final gases}\ (N_2, H_2O, CO_2, CO) \quad (40)$$

The solid HMX (species 1, mass fraction $Y_1$) under high temperature decomposes into fragments (species 2, $Y_2$). The fragments are further decomposed to intermediate gases (species 3, $Y_3$) which are later converted to the final gases (species 4, $Y_4$) through exothermic reactions leading to high temperatures in the hotspot.

# 8 Tables and Figures

| $f_{ignition}^{shape}(AR,\theta)$ | $\theta = 0°$ | $\theta = 15°$ | $\theta = 30°$ | $\theta = 45°$ | $\theta = 60°$ | $\theta = 75°$ | $\theta = 90°$ |
|---|---|---|---|---|---|---|---|
| $AR = 1$ | 0.77 | 0.90 | 2.14 | 1.64 | 1.00 | 0.97 | 0.98 |
| $AR = 4$ | 5.98 | 6.86 | 9.34 | 8.07 | 5.18 | 3.24 | 2.43 |
| $AR = 7$ | 9.83 | 9.67 | 10.41 | 9.22 | 5.81 | 2.72 | 1.13 |
| $AR = 10$ | 11.78 | 10.99 | 10.37 | 9.99 | 5.82 | 2.09 | 0.16 |
| $AR = 13$ | 12.26 | 12.39 | 11.90 | 11.64 | 5.80 | 1.69 | 0.02 |
| $AR = 16$ | 11.63 | 13.48 | 13.54 | 12.69 | 5.34 | 1.16 | 0.00 |
| $AR = 19$ | 11.66 | 14.79 | 15.11 | 13.40 | 5.05 | 0.90 | 0.03 |
| $AR = 22$ | 12.56 | 15.80 | 16.16 | 13.72 | 5.02 | 0.78 | 0.00 |

Table 1 Values of the modifying function for void shape for ignition.

| $f_{growth}^{shape}(AR,\theta)$ | $\theta = 0°$ | $\theta = 15°$ | $\theta = 30°$ | $\theta = 45°$ | $\theta = 60°$ | $\theta = 75°$ | $\theta = 90°$ |
|---|---|---|---|---|---|---|---|
| $AR = 1$ | 0.90 | 1.04 | 1.04 | 1.02 | 1.04 | 1.03 | 1.02 |
| $AR = 4$ | 0.87 | 1.14 | 1.18 | 1.10 | 0.97 | 0.83 | 0.76 |
| $AR = 7$ | 0.92 | 1.26 | 1.29 | 1.09 | 0.71 | 0.39 | 0.29 |
| $AR = 10$ | 0.93 | 1.32 | 1.37 | 1.12 | 0.61 | 0.18 | 0.03 |
| $AR = 13$ | 0.98 | 1.37 | 1.43 | 1.17 | 0.61 | 0.16 | 0.00 |
| $AR = 16$ | 1.24 | 1.53 | 1.54 | 1.24 | 0.61 | 0.19 | 0.00 |
| $AR = 19$ | 1.49 | 1.67 | 1.65 | 1.27 | 0.56 | 0.21 | 0.01 |
| $AR = 22$ | 1.66 | 1.77 | 1.76 | 1.33 | 0.55 | 0.22 | 0.00 |

Table 2 Values of the modifying function for void shape for growth.

| $\phi$ (%) | *Minimum Inter-Void Distance* ($\mu m$) |
|---|---|
| 2 | 45 |
| 5 | 25 |
| 7 | 20 |
| 10 | 15 |
| 12 | 10 |
| 15 | 7.5 |
| 20 | 5 |

Table 3: The minimum inter-void distance specified for a given void volume fraction $\phi$.

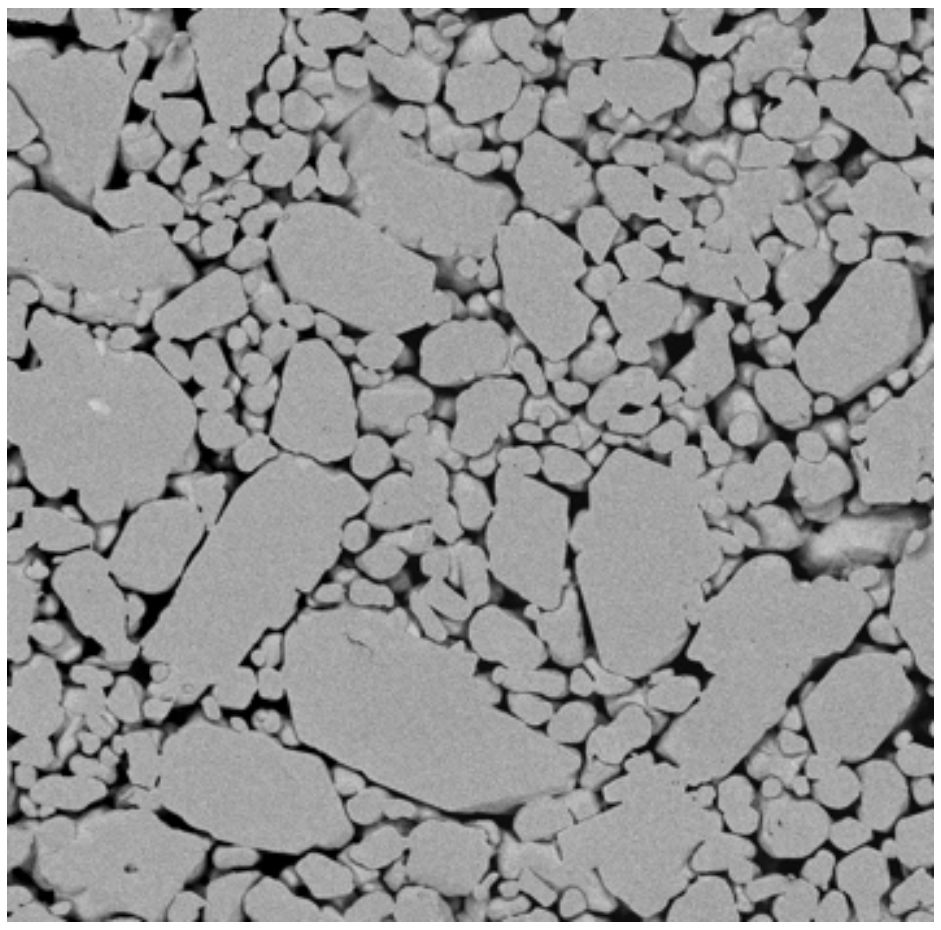

Figure 1: SEM image [6] of a Class V HMX sample of size 20.0 $\mu m$ x 20.0 $\mu m$. The black regions indicate the voids and the grey represent solid HMX.

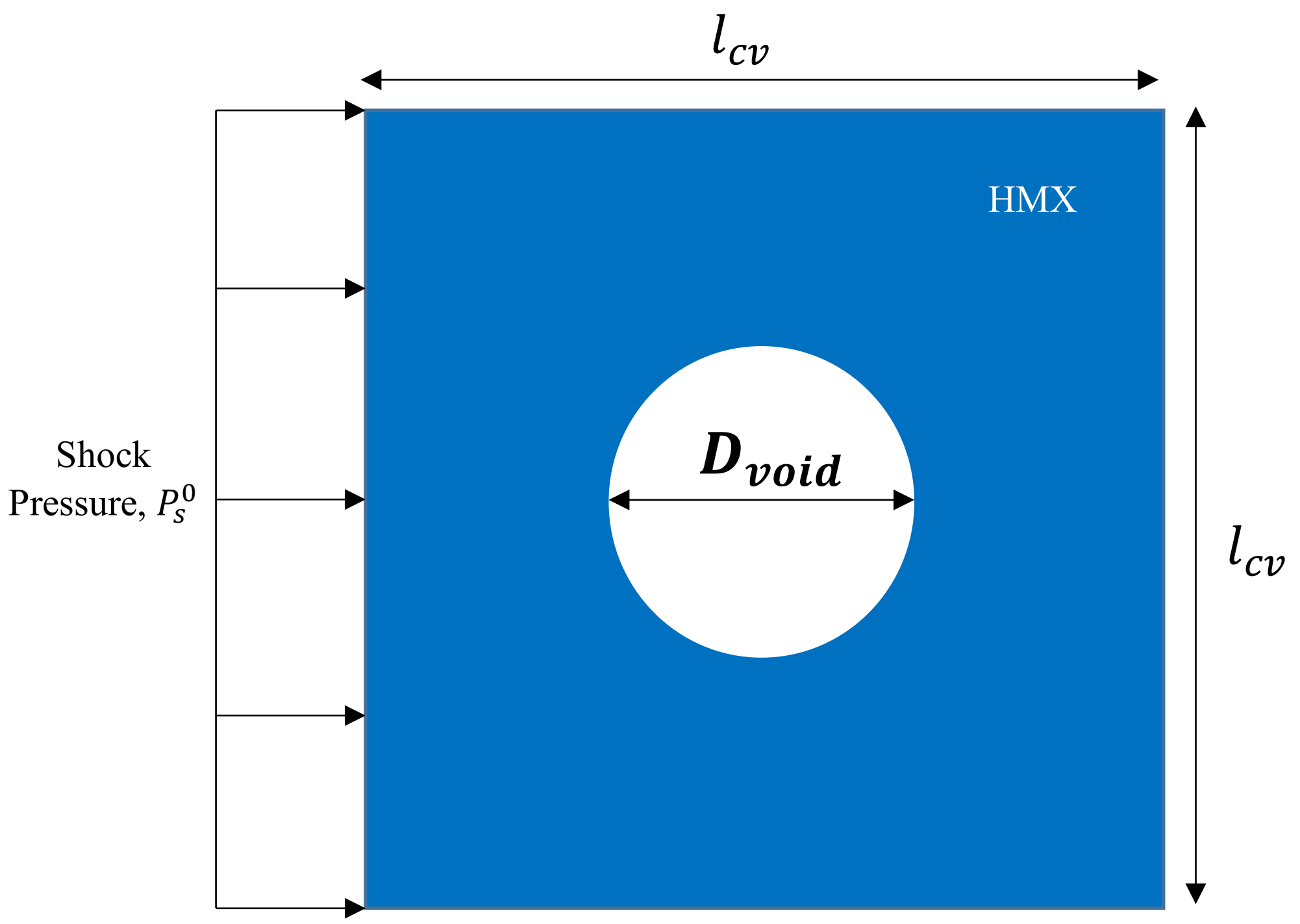

Figure 2: The numerical setup for performing high-fidelity mesoscale simulations of reactive void-collapse to construct the surrogate $\dot{F}_{\mathrm{ignition}}^{\mathrm{scv}}$ and $\dot{F}_{\mathrm{growth}}^{\mathrm{scv}}$. A single cylindrical void with diameter $D_{\mathrm{void}}$ is subjected to a shock pressure $P_{\mathrm{s}}$, with $\tau_{\mathrm{s}}$ denoting the duration of the loading. The load is applied from the west side of the domain boundary, as shown in the figure. Shock load is applied as a rectangular pulse from the left side of the domain. Zero gradient boundary conditions are applied for the other three boundaries. Control volume length $l_{cv} = 30\mu m$.

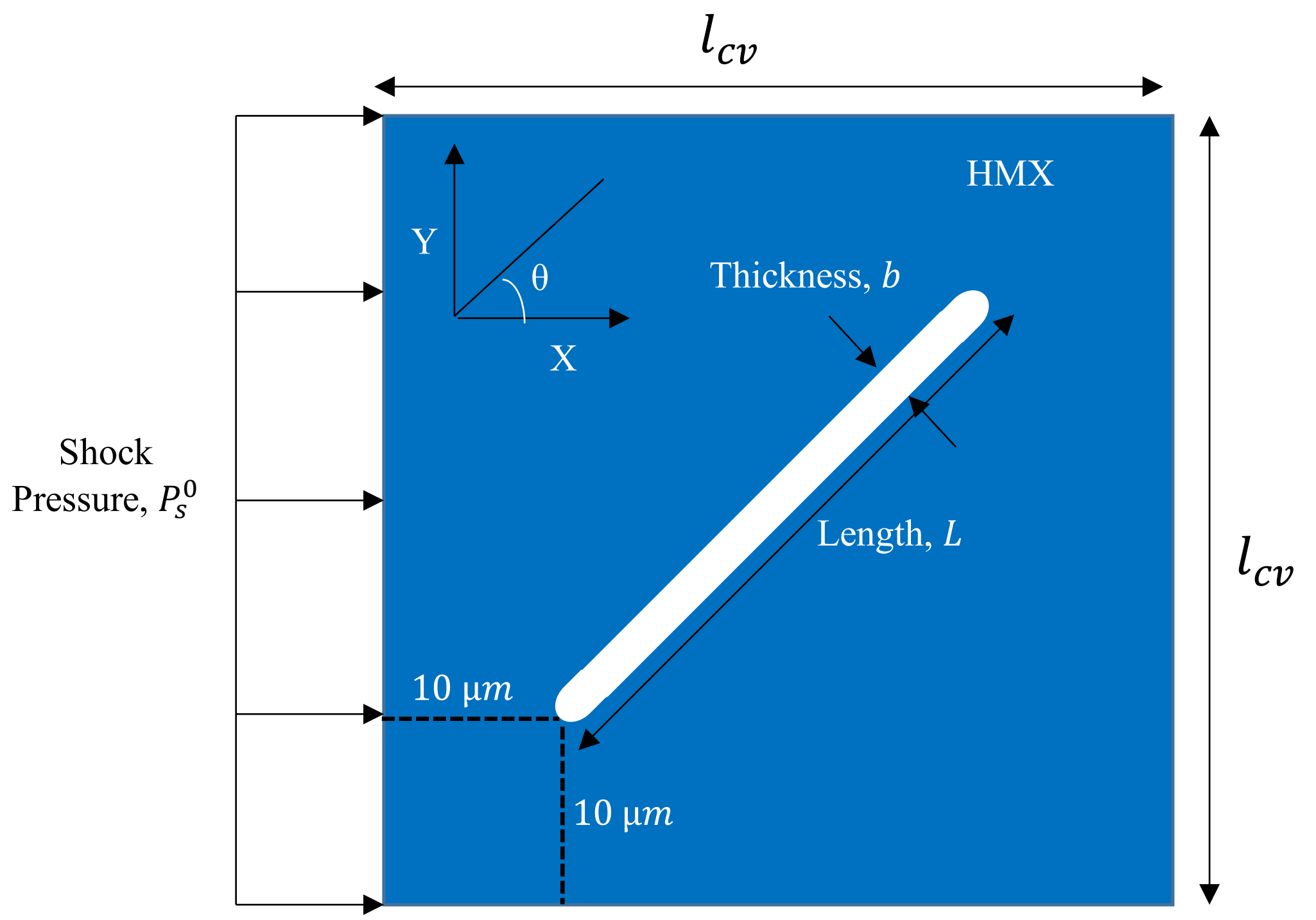

Figure 3: Numerical set up for void collapse analysis of elongated void of length, $L$ and thickness, $b$ oriented at angle of $\theta$ with respect to the reference loading condition. Shock load is applied as a rectangular pulse from the left side of the domain. Zero gradient boundary conditions are applied for the other three boundaries. Control volume length $l_{cv} = 50\ \mu m$.

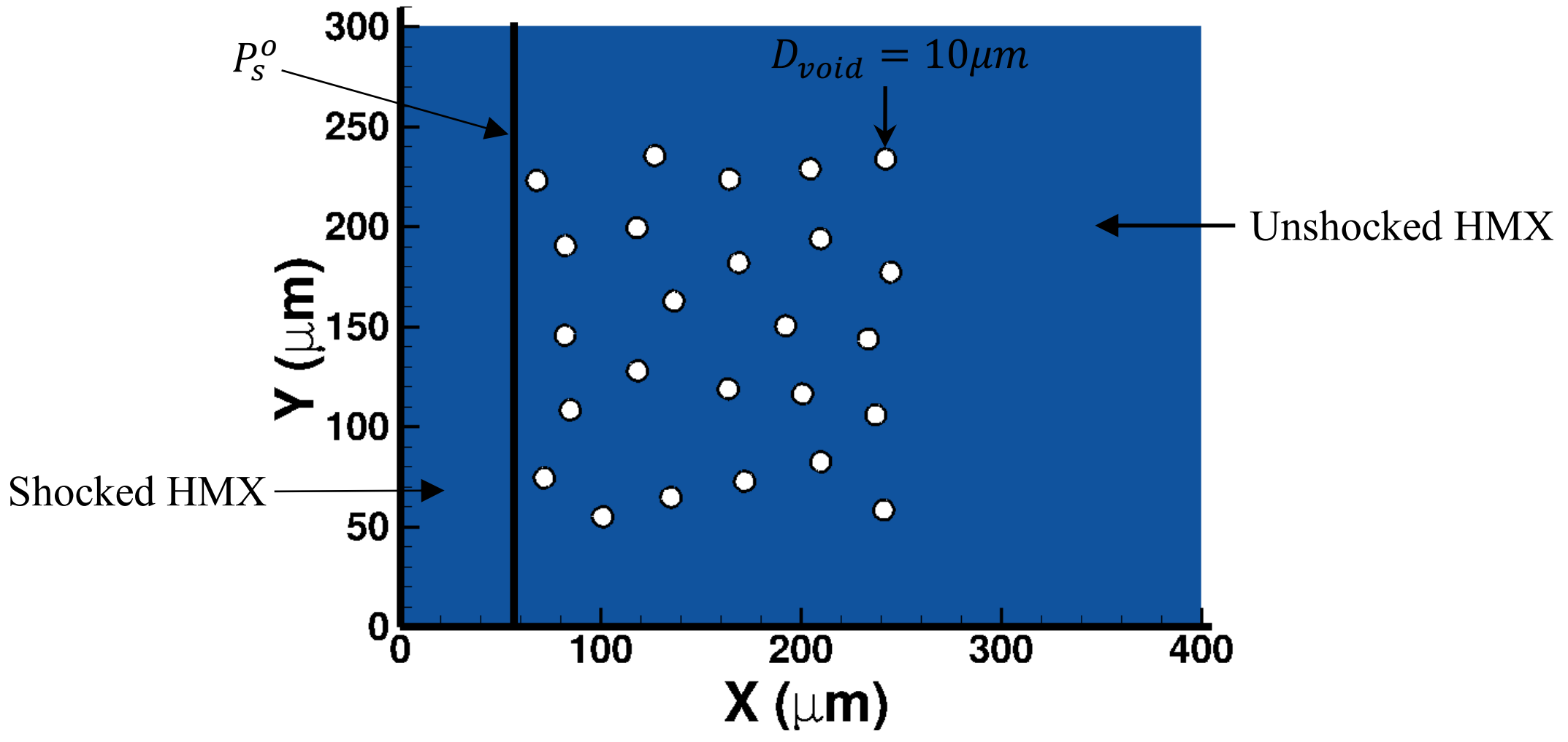

Figure 4: Numerical setup for simulations of a field of cylindrical voids at the reference loading condition.

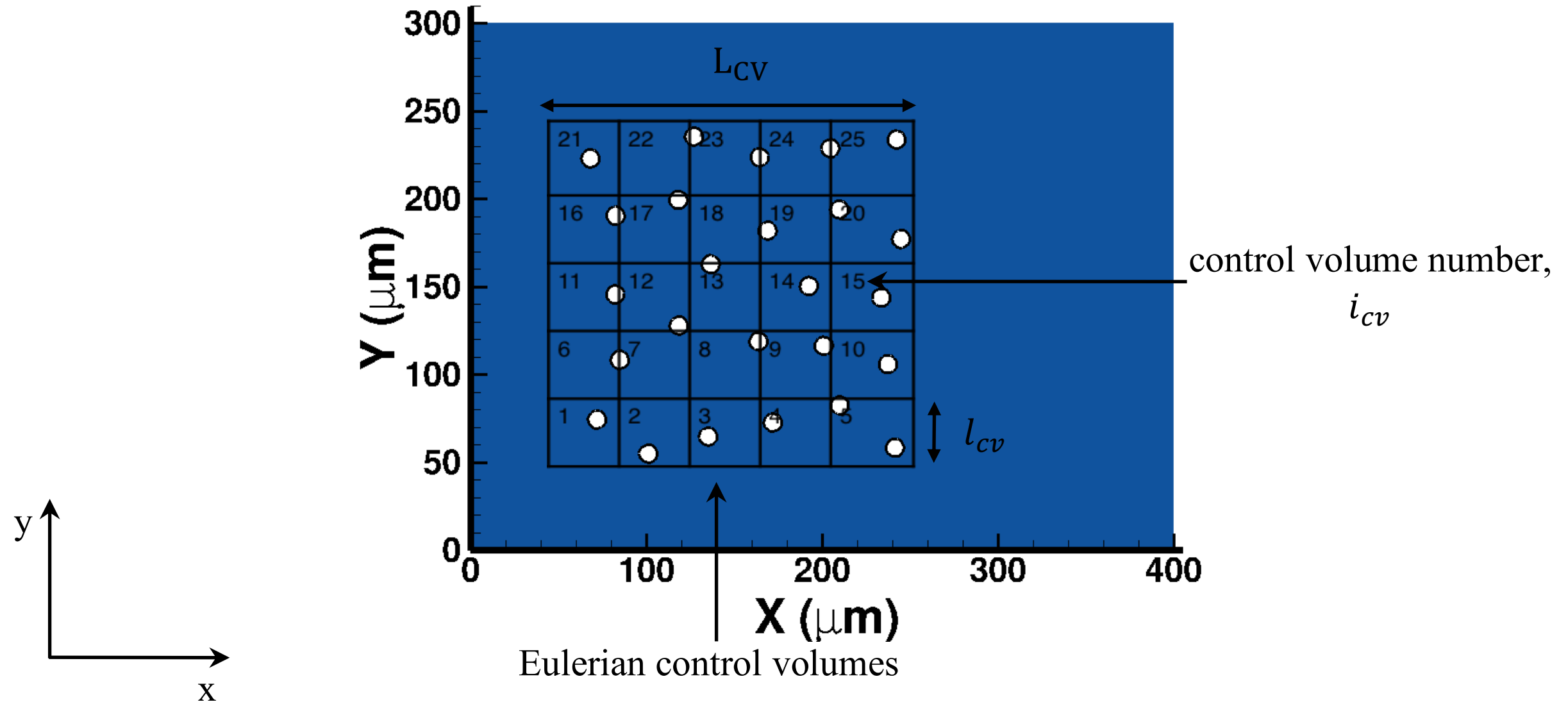

Figure 5: Illustration of a square Eulerian control volume of length $L_{CV}$, moving with velocity $V_{CV}$. The larger control volume is subdivided in to a total number of $N_{cv}$ smaller control volumes $cv$ of length $l_{cv}$. In the above setup there are 25 voids with a volume fraction of $\phi = 5$ %, and $N_{cv} = 25$.

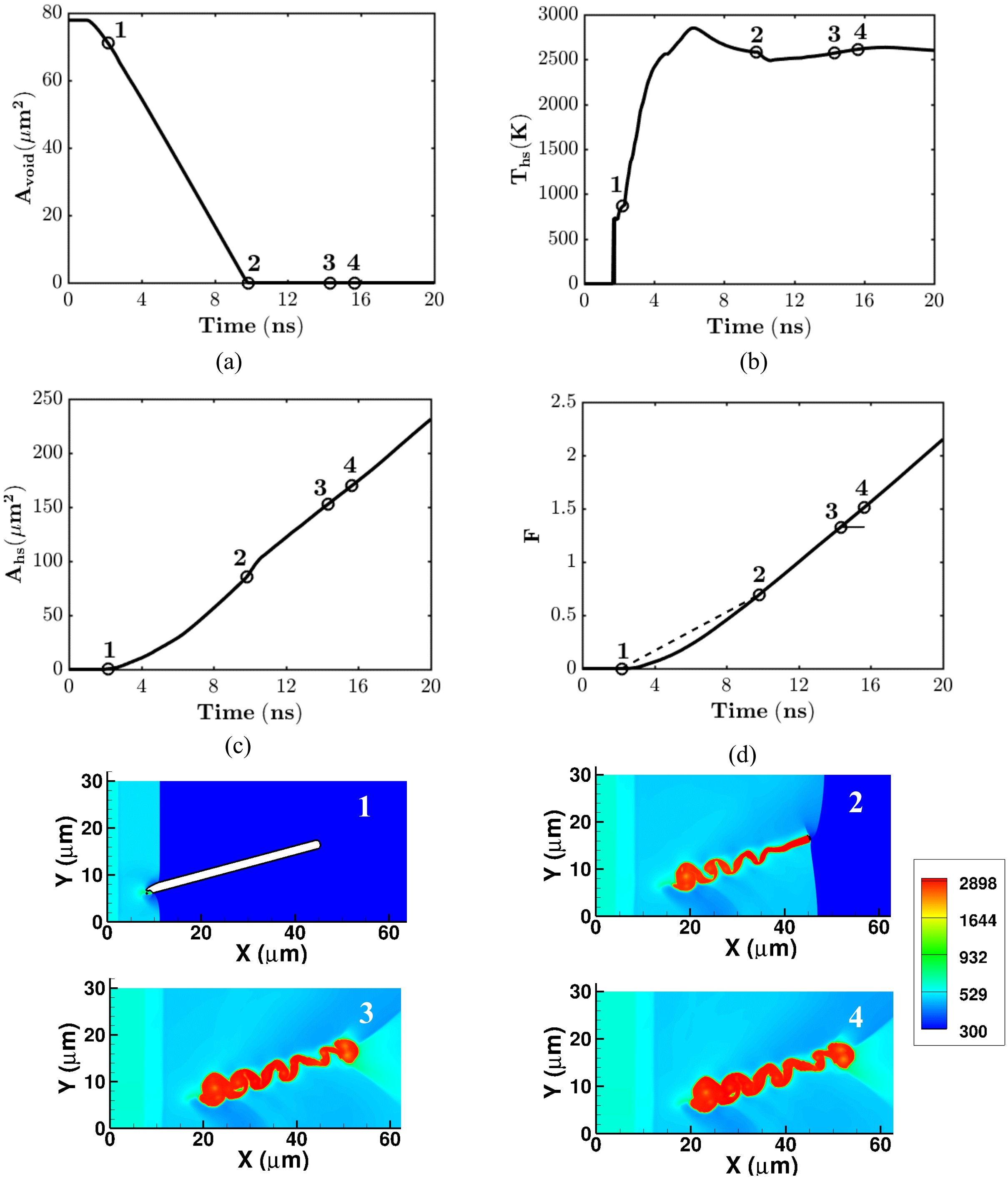

Figure 6: Simulation results for the void collapse of an elongated void of length 41.536 $\mu m$ and thickness 1.89 $\mu m$ with aspect ratio, $AR = 22$ oriented at an angle of $\theta = 15°$ with respect to the reference loading condition. The figure shows the time variation of (a) void area, (b) hotspot average temperature, (c) hotspot area, and (d) the products mass fraction. Four different instances of void collapse are marked at each figure with numbers 1 – 4 with the corresponding temperature contours.

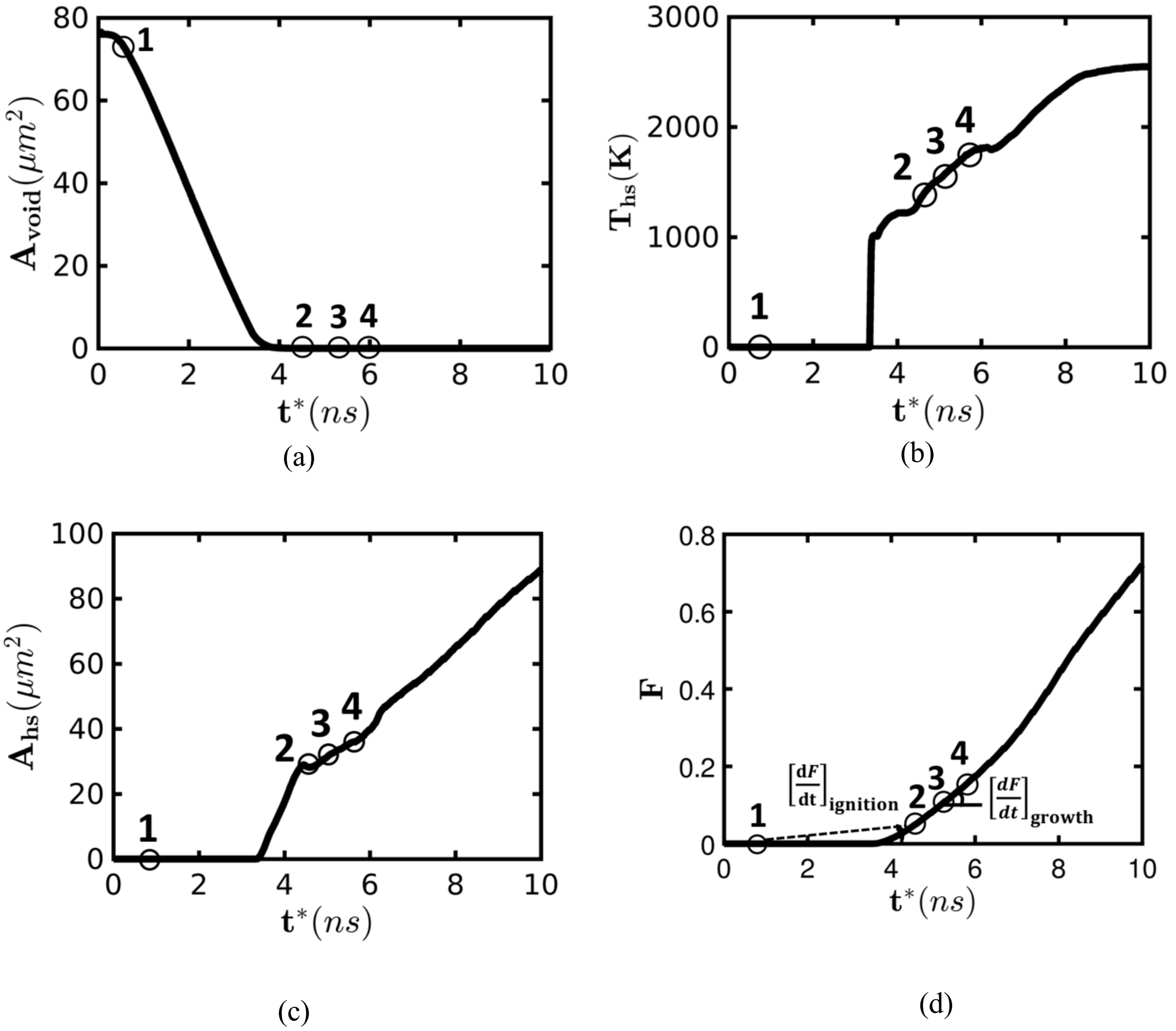

Figure 7: Ignition and growth stages for $\phi = 5\%$ and control volume number $i_{cv} = 13$. **1** and **2** indicate the beginning and end of the ignition phase, respectively. **3** and **4** indicate the start and end of the growth phase, respectively. (a) Variation of the average $A_{void}$ with respect to $t^*$, indicating the void area in the control volume. (b) Variation of $T_{hs}$ with $t^*$indicating the average hotspot temperature in the control volume. (c) Variation of the hotspot area $A_{hs}$ with $t^*$, and (d) variation of $F$ with $t^*$ within the control volume.

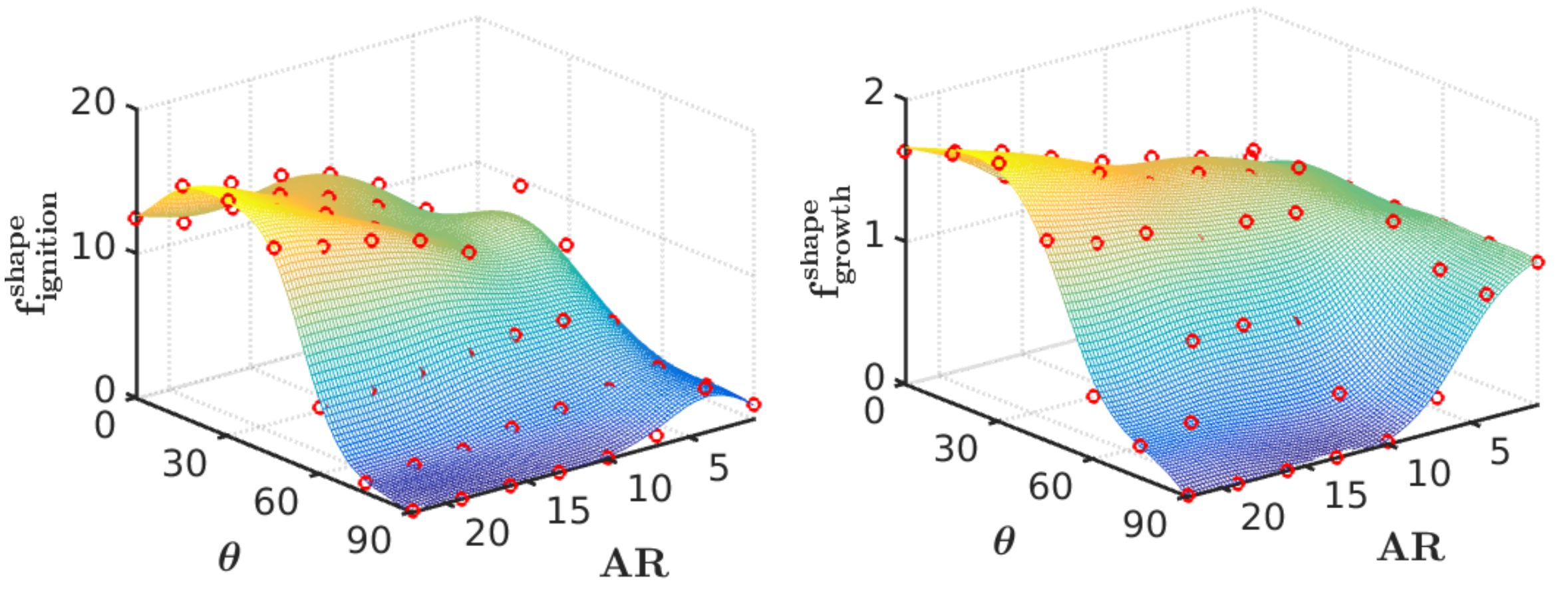

(a) Surrogate Model of $f_{ignition}^{shape}(AR, \theta)$ (b) Surrogate Model of $f_{growth}^{shape}(AR, \theta)$

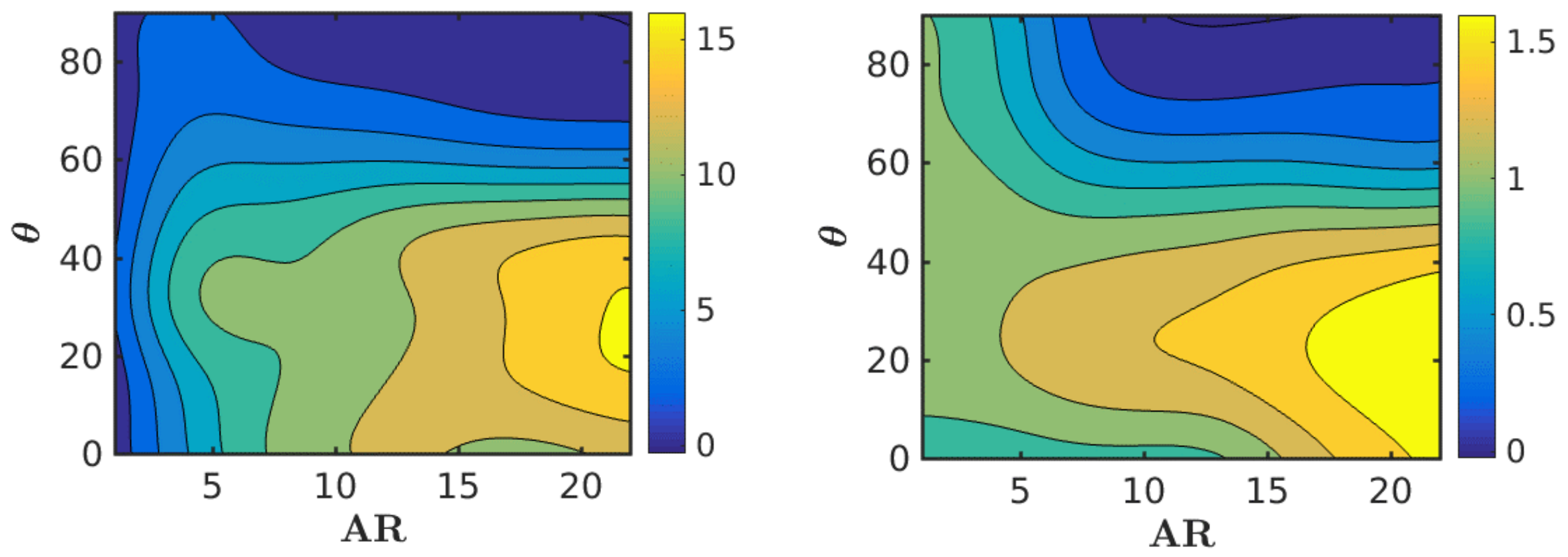

(c) Contour plot of $f_{ignition}^{shape}(AR, \theta)$ surrogate (d) Contour plot of $f_{growth}^{shape}(AR, \theta)$ surrogate

Figure 8: Surrogate models for the modifier functions $f_{ignition}^{shape}(AR, \theta)$ and $f_{growth}^{shape}(AR, \theta)$ for $AR = \{1, 4, 7, 10, 13, 16, 19, 22\}$ and $\theta = \{0°, 15°, 30°, 45°, 60°, 75°, 90\}$. The surrogate models are obtained for the reference loading condition $P_s^0$ and $\tau_s^0$.

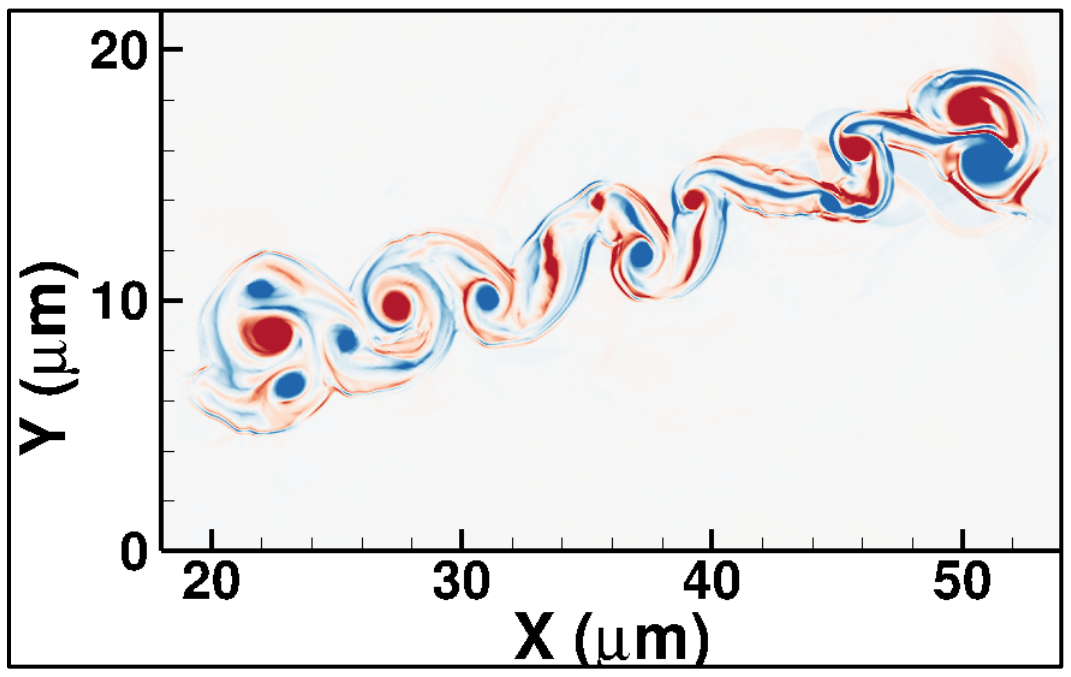

(a) Vorticity contours at $t = 14.3\ ns$ for the elongated void

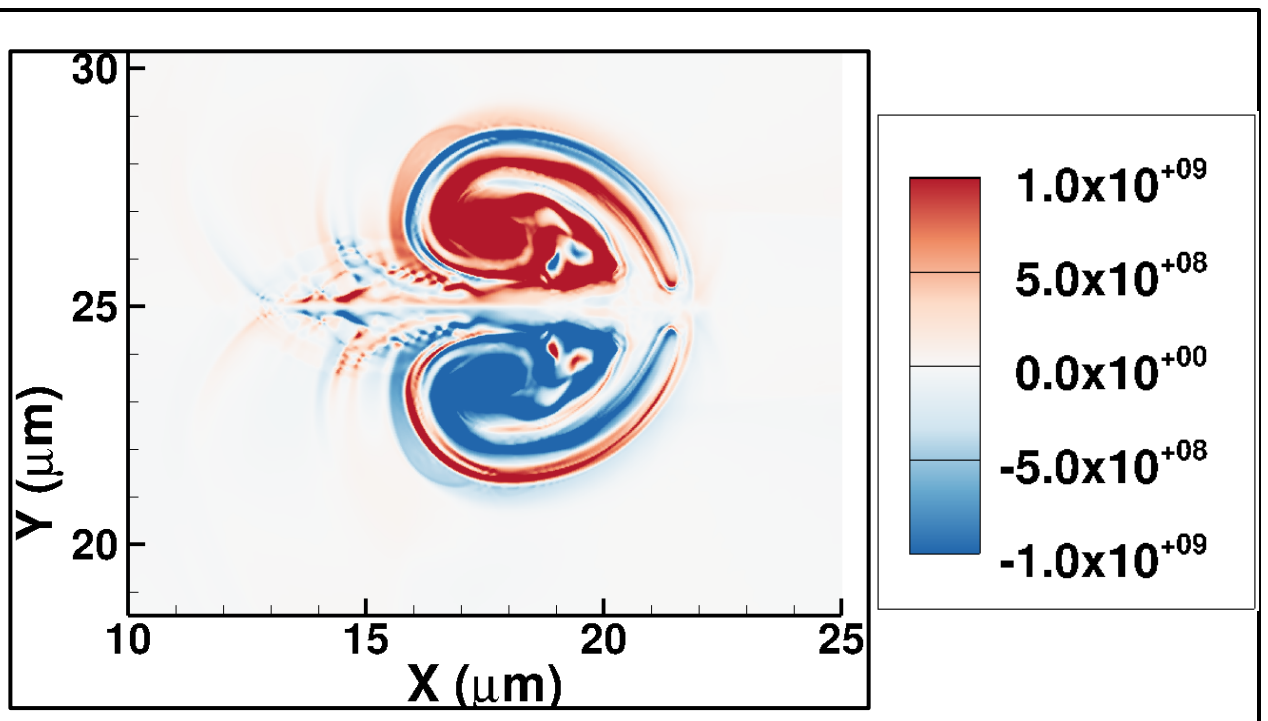

(b) Vorticity contours at $t = 7.5\ ns$ for the cylindrical void

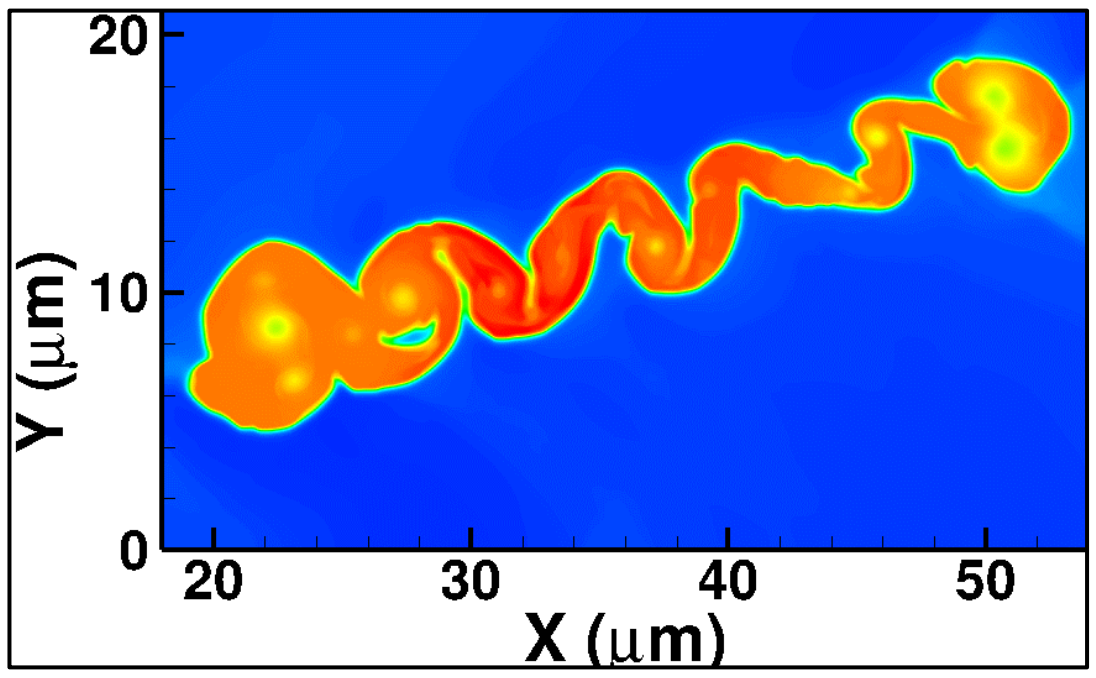

(c) Temperature (K) contours at $t = 14.3\ ns$ for the elongated void

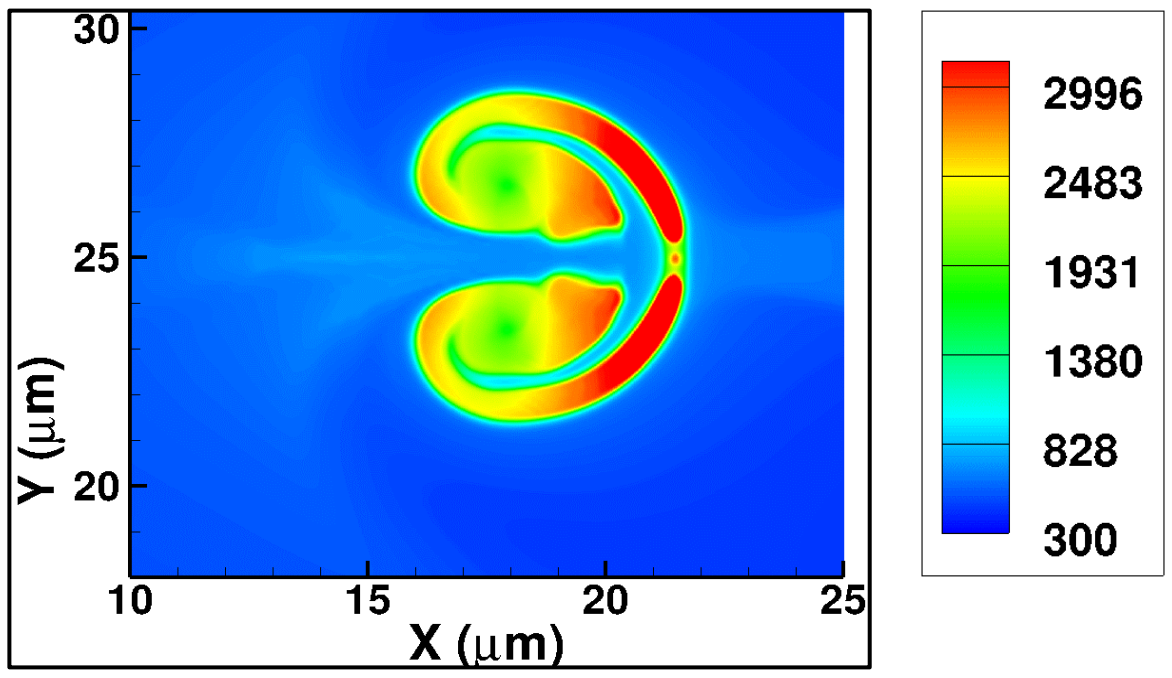

(d) Temperature (K) contours at $t = 7.5\ ns$ for the cylindrical void

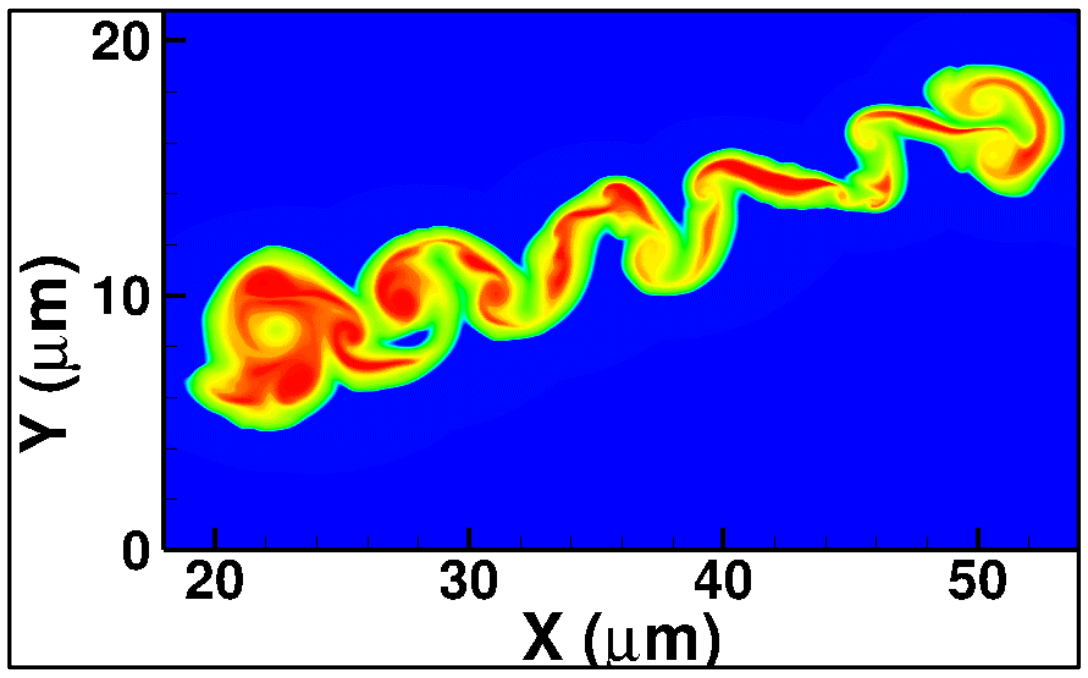

(e) Reaction products mass fraction at $t = 14.3\ ns$ for the elongated void

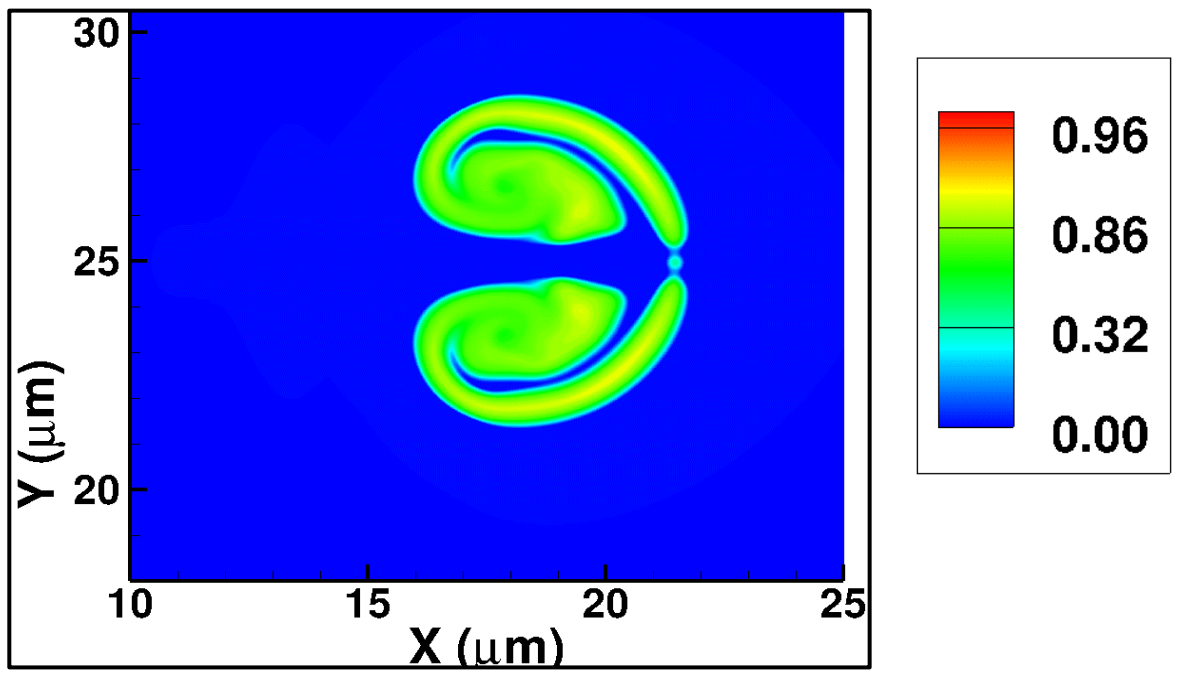

(d) Reaction products mass fraction at $t = 7.5\ ns$ for the cylindrical void

Figure 9: Vorticity, temperature and final reaction products mass fraction contour plots for the elongated void and the cylindrical void of same void area for the reference loading condition. The elongated void area is equivalent to $D_{void}^{0}$ and oriented at an angle of 15° with respect to the incident shock.

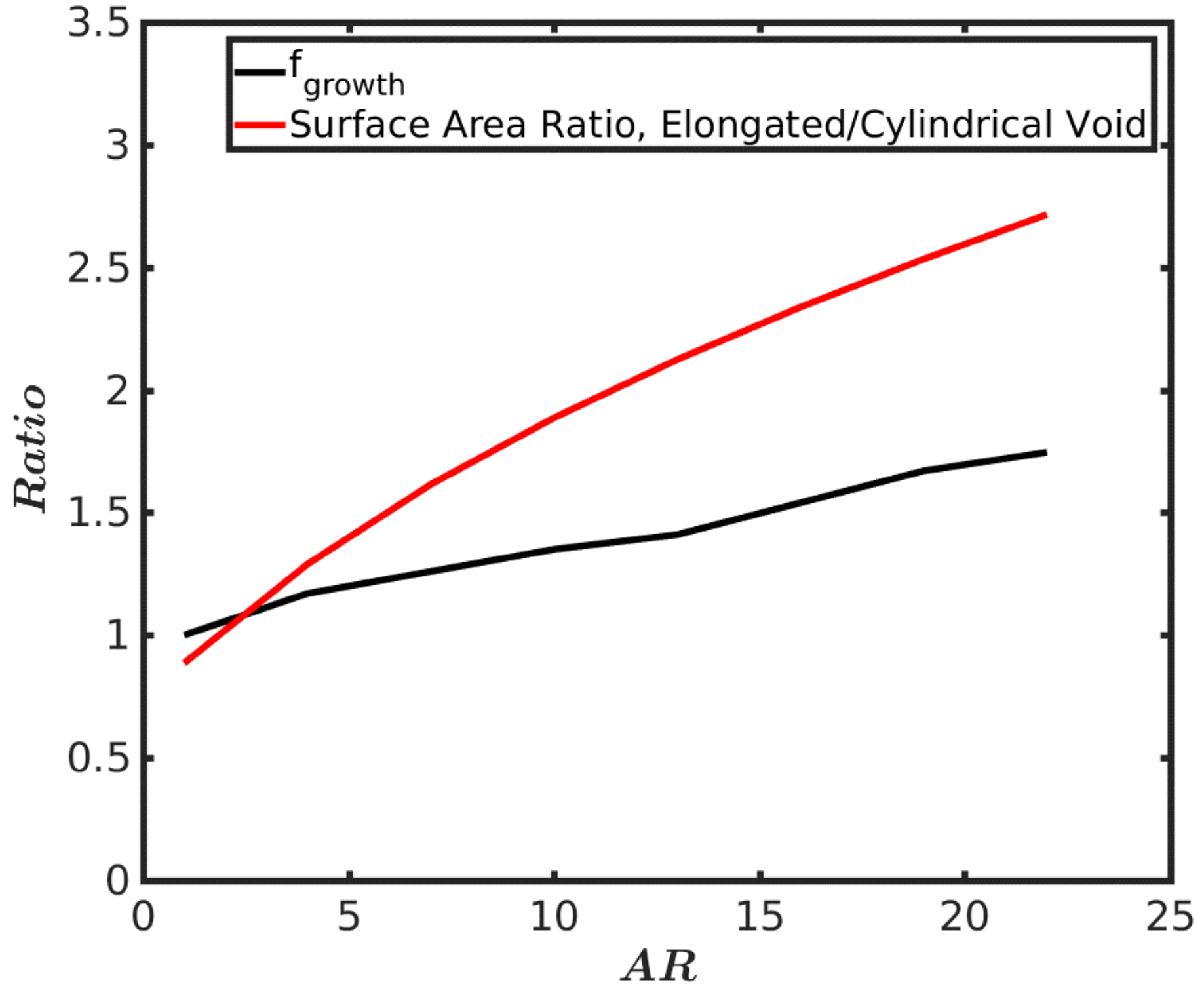

Figure 10: Variation of the surface area ratio of elongated and cylindrical voids and $f_{growth}^{shape}(AR, \theta = 15°)$ with $AR$. The elongated void area is equivalent to $D_{void}^{0}$. The amplification of the hotspot growth rates for elongated voids (black curve) is weaker than expected from the ratio of surface areas of the elongated and cylindrical voids (red curve).

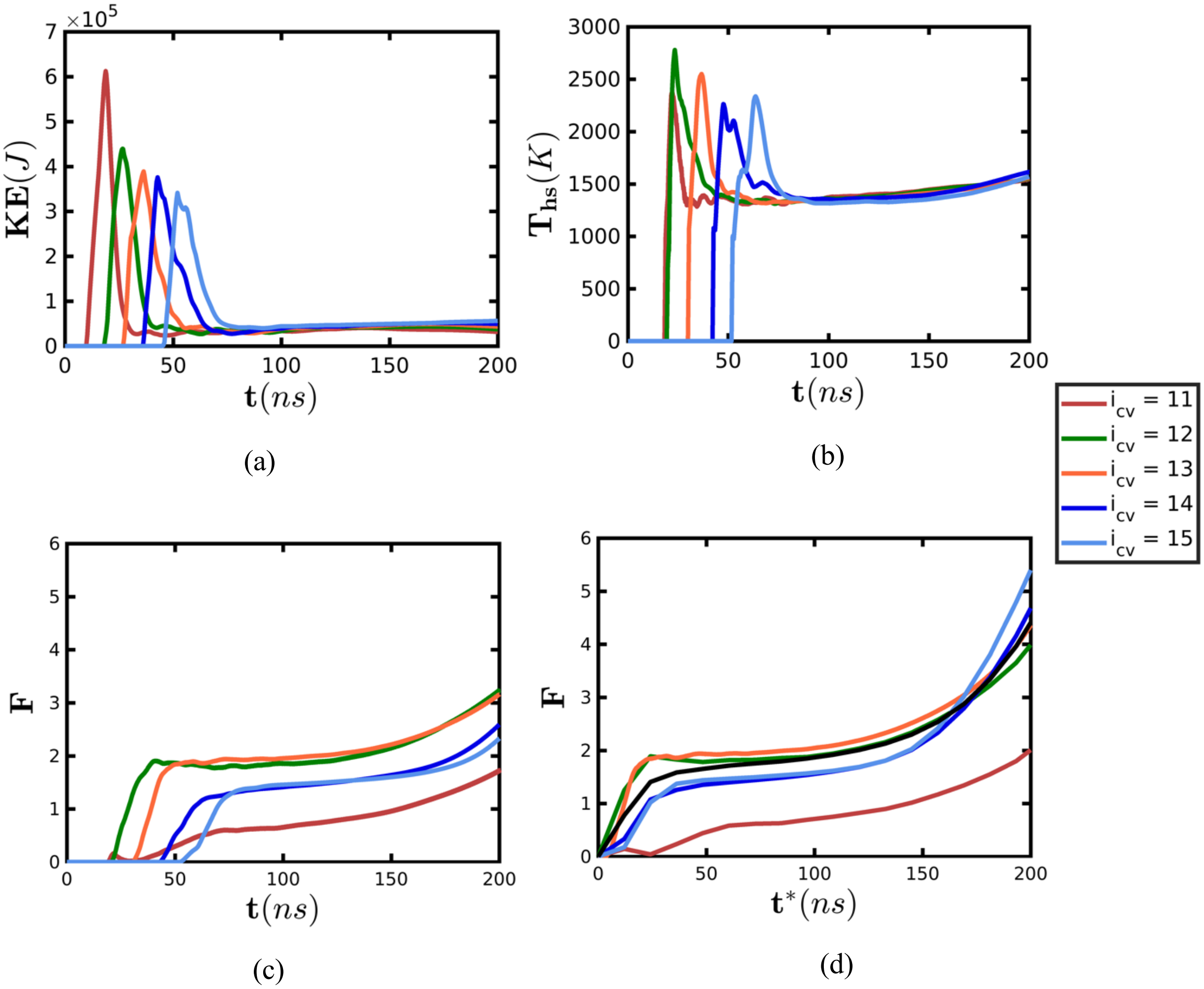

Figure 11: Results for $\phi = 5\%$ with random arrangement, where $i_{cv}$ is the control volume number. (a) Temporal variation of kinetic energy $KE_{cv}$ (b) temporal variation of hotspot temperature $T_{hs}$, (c) temporal variation of $F$, and (d) variation of $F$ with $t^*$, where $t^*$ indicates time has been shifted based on the kinetic energy rise for control volume number $i_{cv}$.

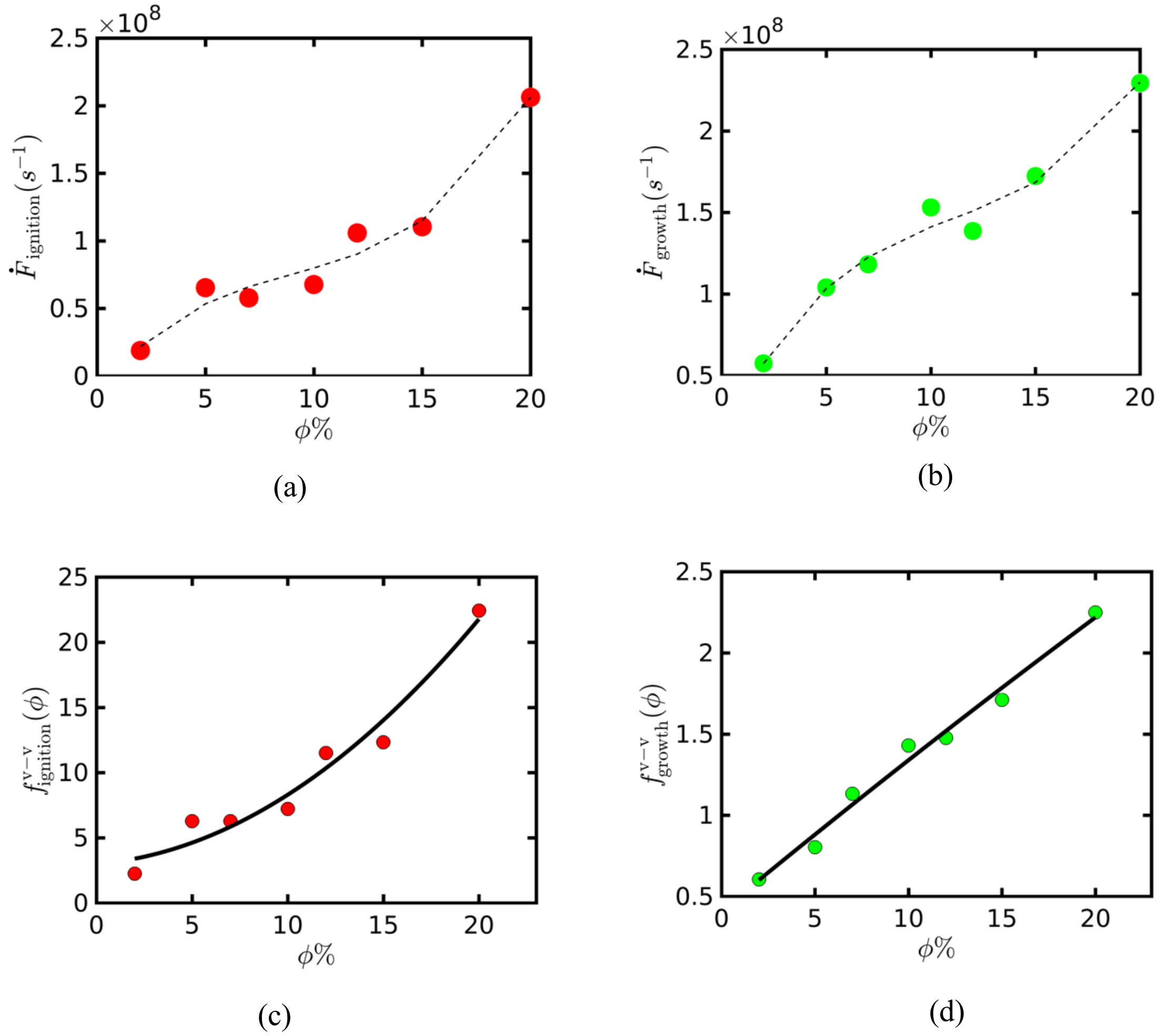

Figure 12: Effect of volume fraction $\phi$ on (a) $\dot{F}_{ignition}$, (b) $\dot{F}_{growth}$, (c) $f^{v-v}_{ignition}(\phi)$, and (d) $f^{v-v}_{growth}(\phi)$.

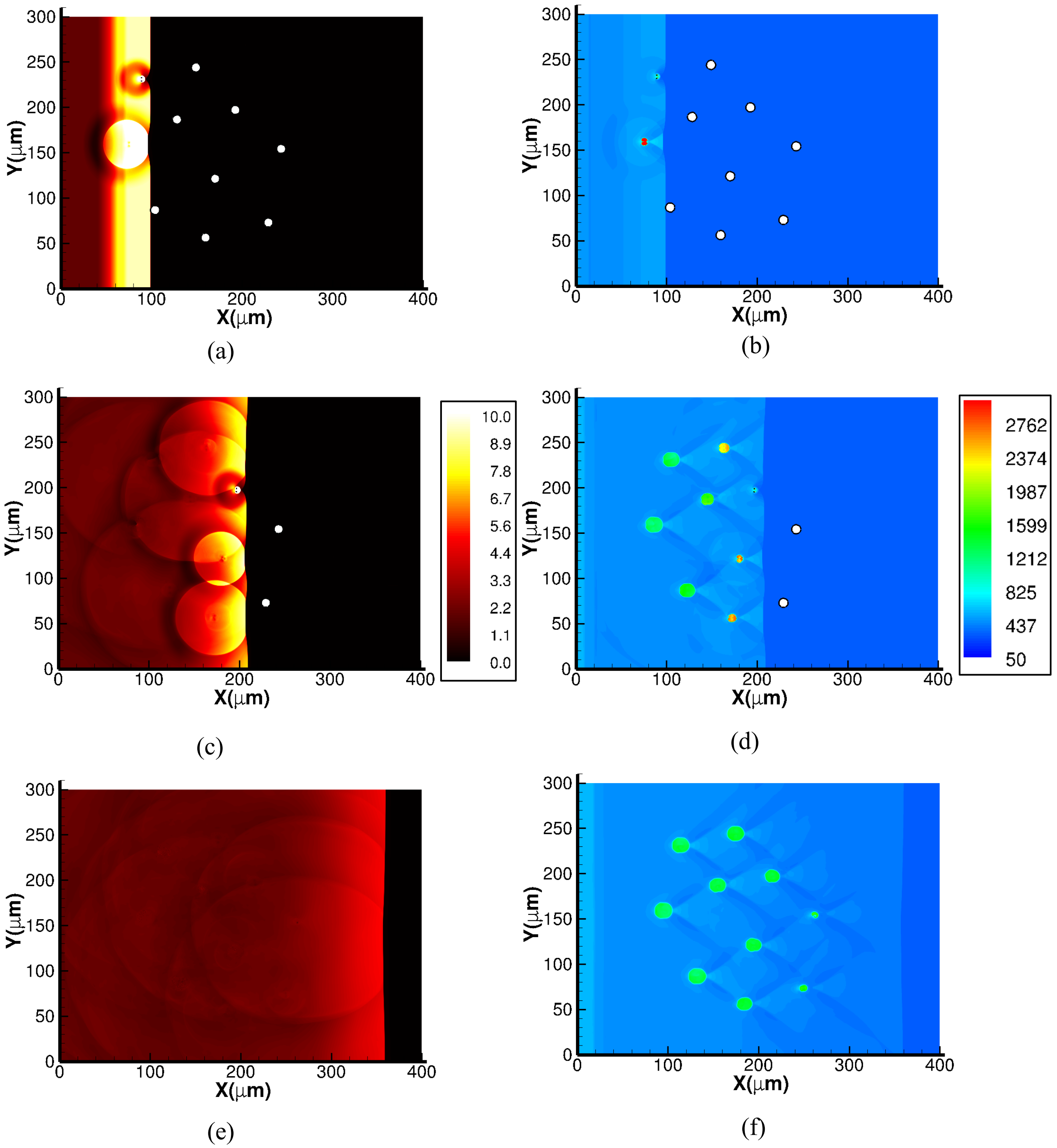

Figure 13: Reactive simulation at reference loading condition for $\phi = 2\%$ with pressure (in GPa) contour plots at time instances (a) $t = 19.8\ ns$, (c) $t = 42.3\ ns$, and (e) $t = 76.1\ ns$. Similarly, the temperature (in K) contours are shown at time instances (b) $t = 19.8\ ns$, (d) $t = 42.3\ ns$, and (f) $t = 76.1\ ns$.

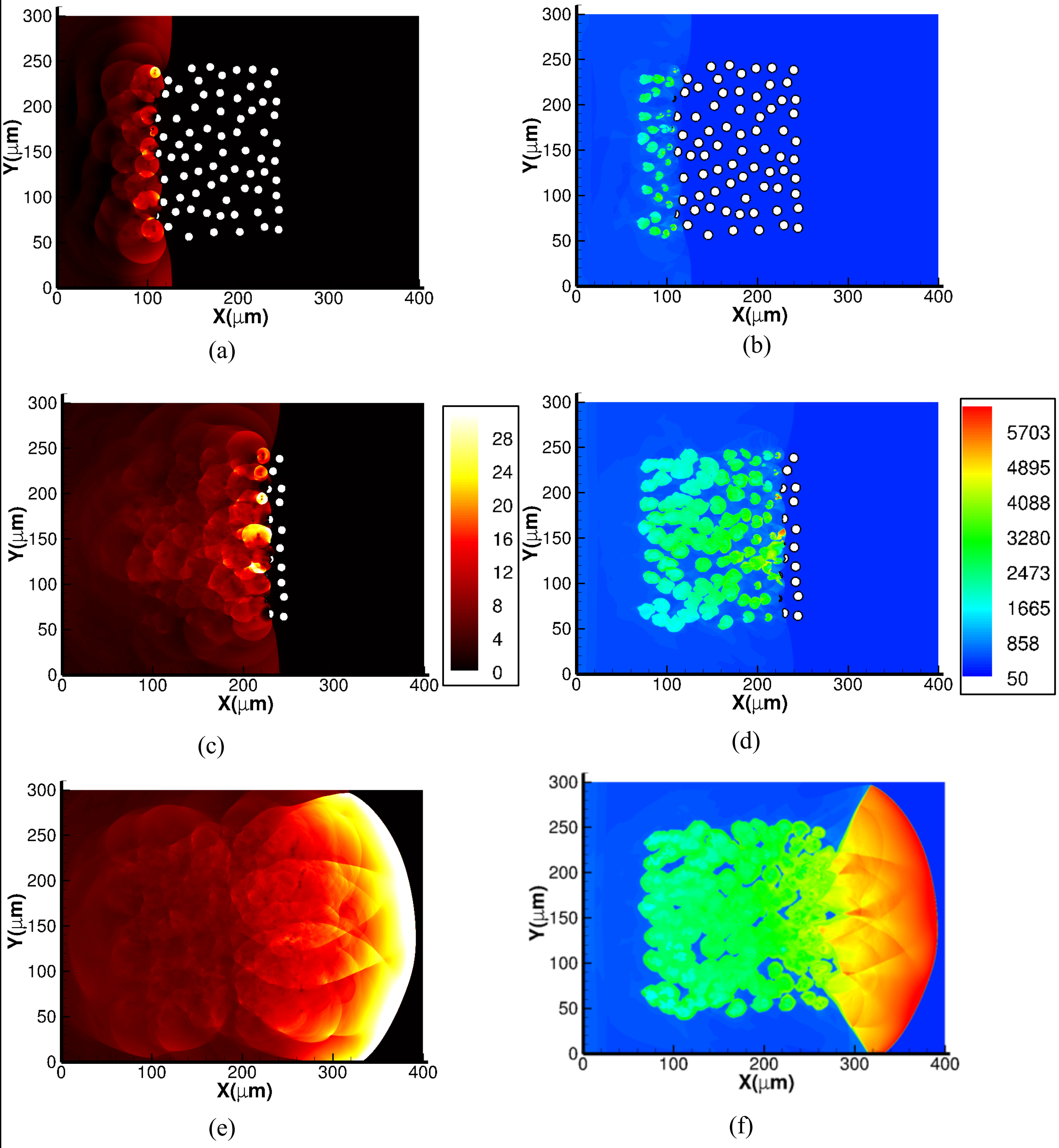

Figure 14: Reactive simulation at reference loading condition for $\phi = 20\%$ with pressure (in GPa) contour plots at time instances (a) $t = 25.3\ ns$, (c) $t = 49.7\ ns$ and (e) $t = 68.2\ ns$. Similarly, the temperature (K) contours are shown at time instances (b) $t = 25.3\ ns$, (d) $t = 49.7\ ns$ and (f) $t = 68.2\ ns$.

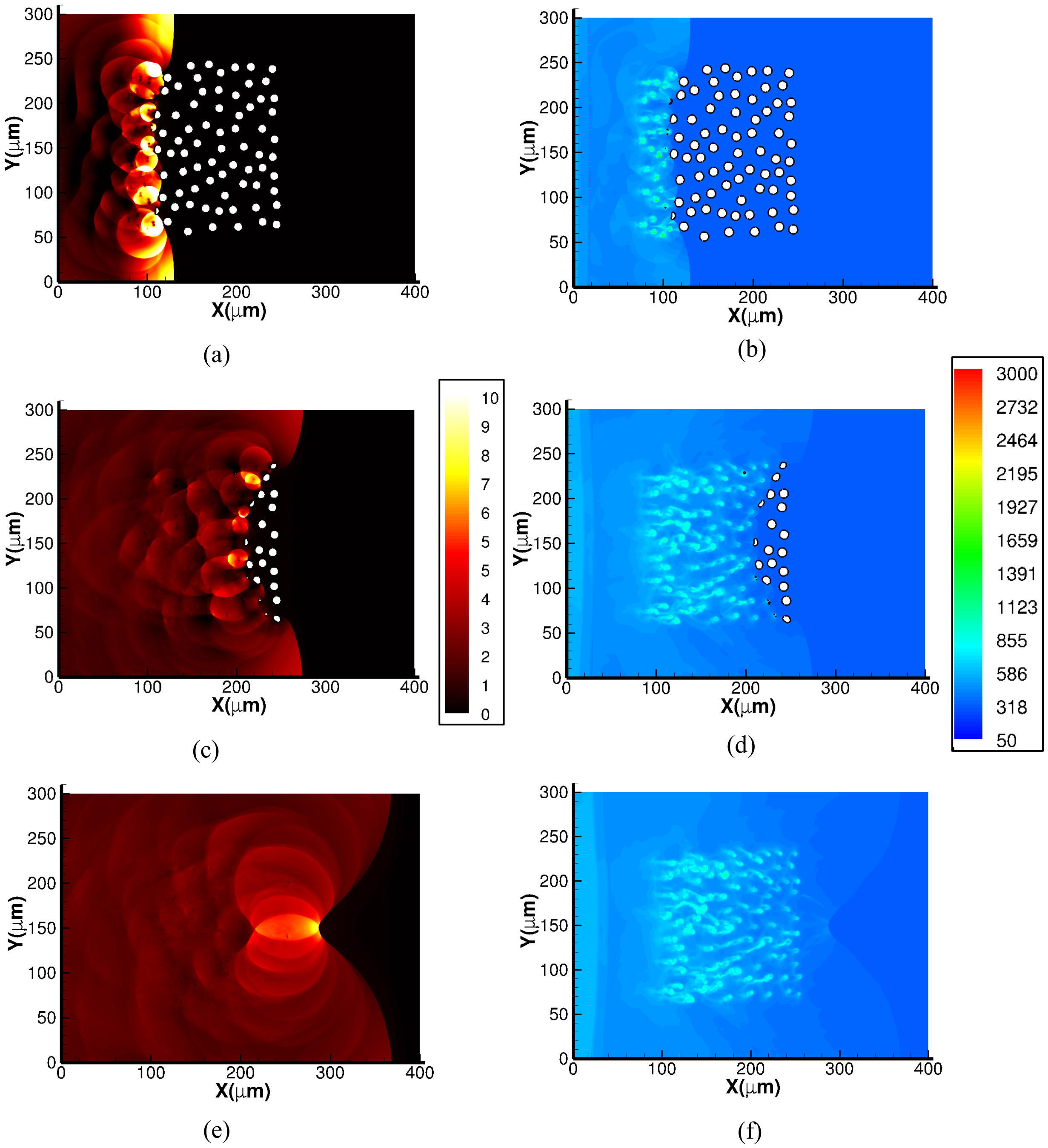

Figure 15: Inert simulation for reference loading condition at $\phi = 2\%$ with pressure (in GPa) contour plots at time instances (a) $t = 26\ ns$, (c) $t = 58.1\ ns$ and (e) $t = 82.1\ ns$. Similarly, the temperature (K) contours are shown at time instances (b) $t = 26\ ns$, (d) $t = 58.1\ ns$ and (f) $t = 82.1\ ns$.

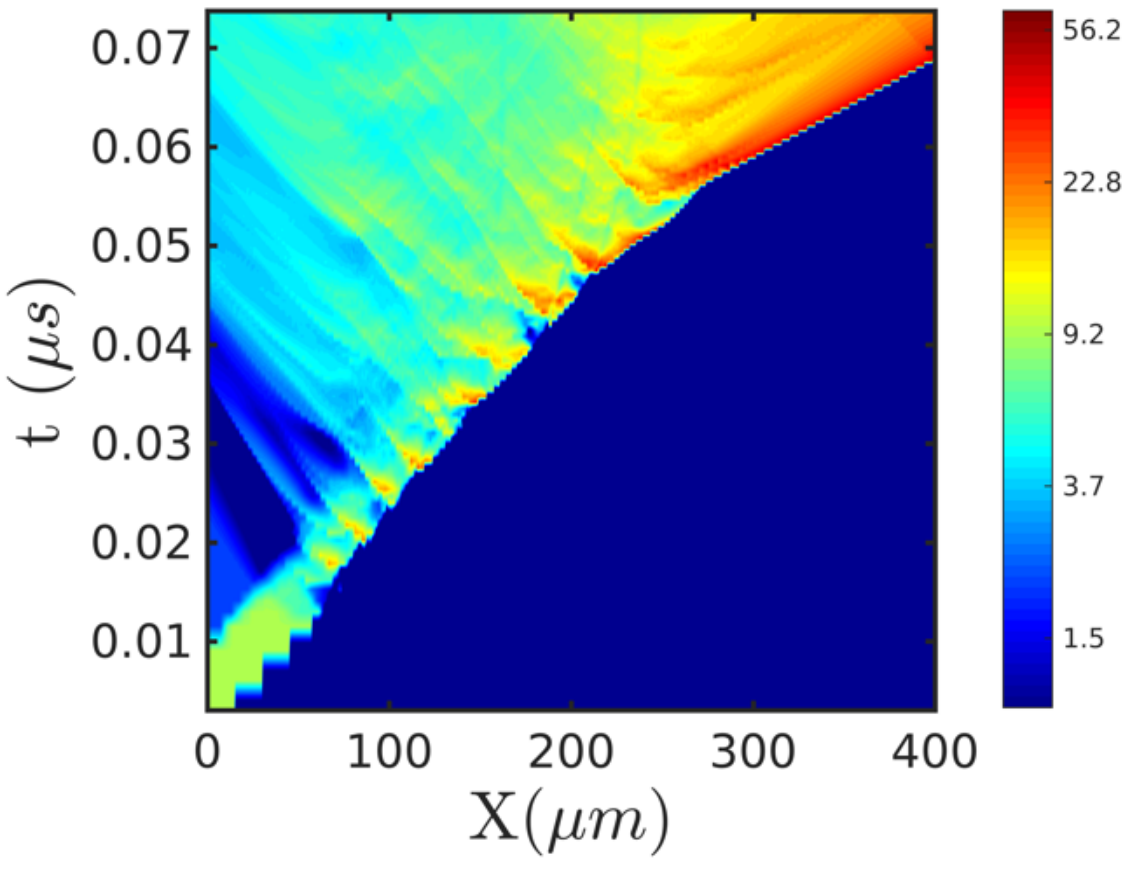

(a)

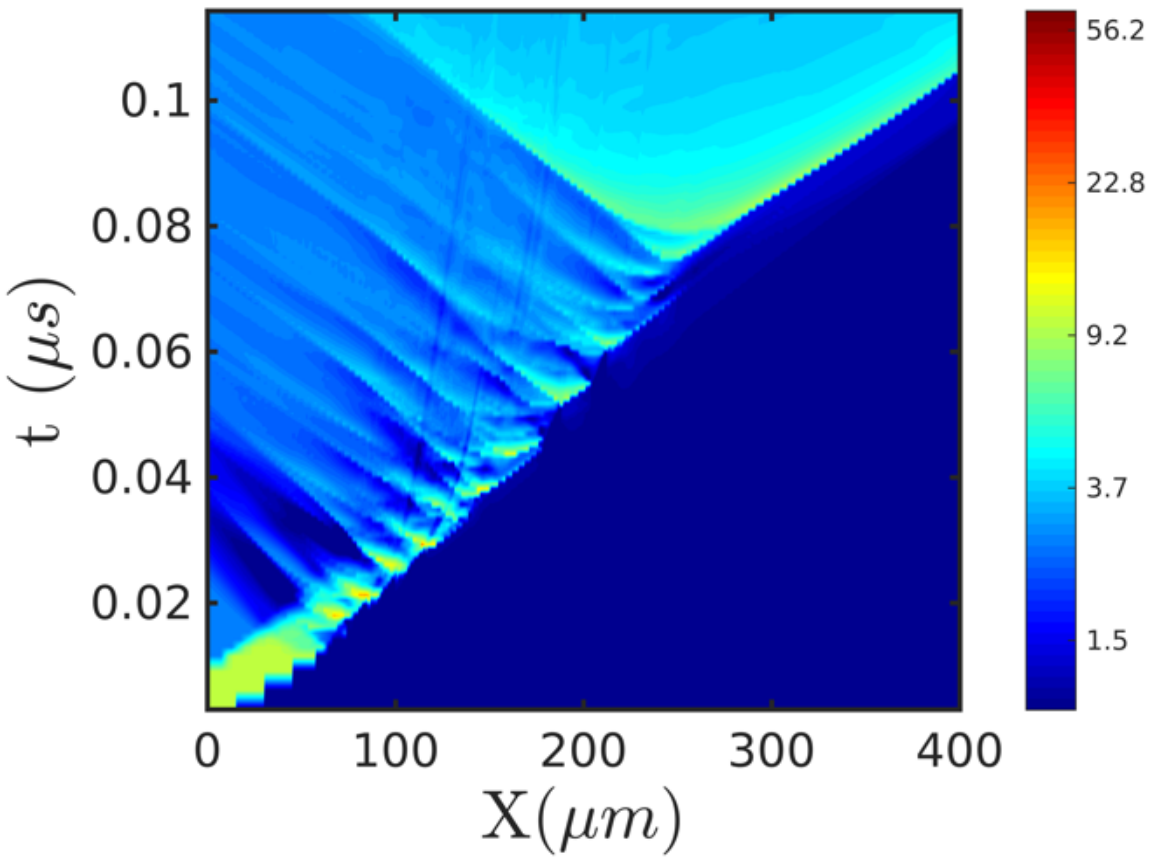

(b)

Figure 16: *x*-*t* plots for the pressure (in GPa) at the reference loading conditions for $\phi = 20\%$ for the (a) reactive simulation and (b) the inert simulation. The *x*-*t* plots are obtained for $Y = 150\mu m$.

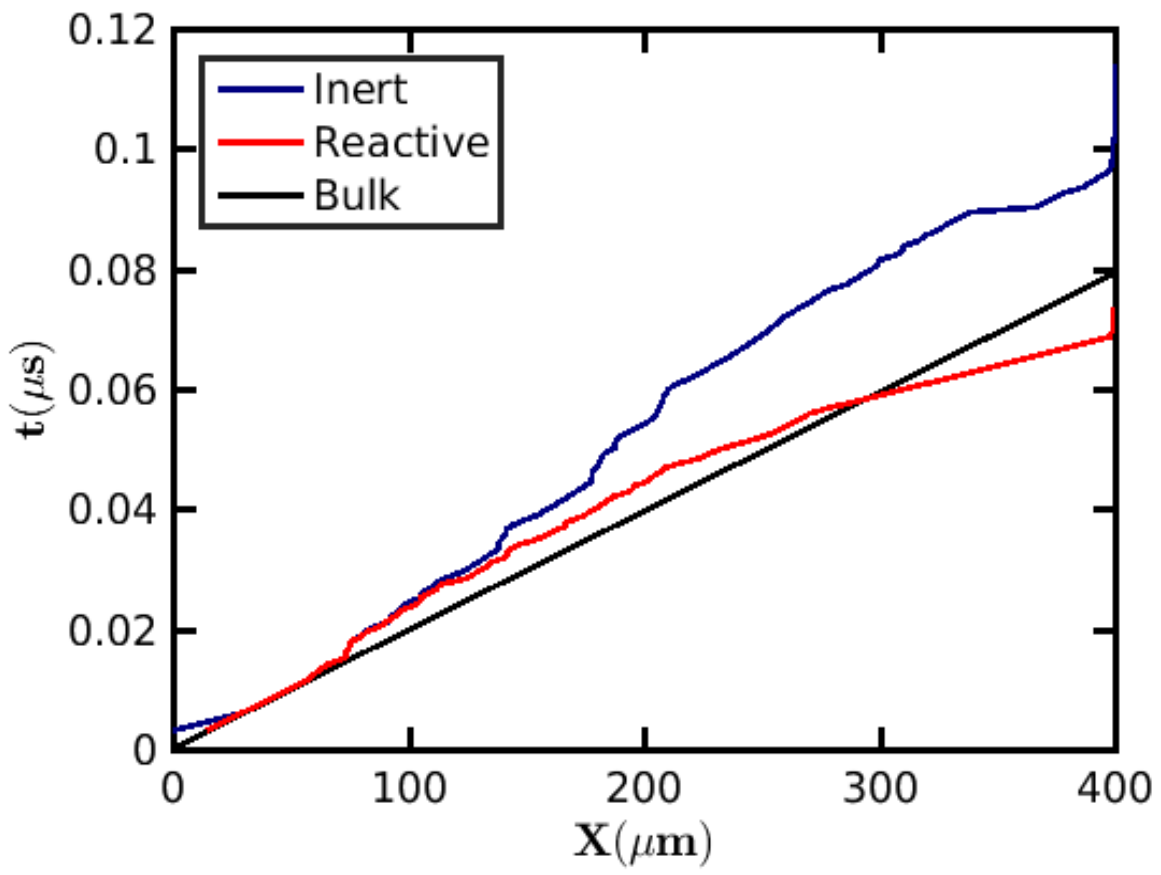

Figure 17: Comparison of the shock front in the *x*-*t* plot at the reference loading condition for the inert, reactive, and bulk simulation, where bulk implies the shock passes through bulk HMX sample without any voids.